\documentclass[aip,cha,reprint,amsmath,amssymb]{revtex4-1}
\usepackage{graphicx}
\usepackage{epsf}
\begin{document}

\title{ Dynamic Phase Transition from Localized to Spatiotemporal Chaos in 
Coupled Circle Map with Feedback.}

\author{Abhijeet R. Sonawane} 

\affiliation{
 Center for Modeling and Simulation,
University of Pune, 
 Pune-411 007,  India}
\author{Prashant M. Gade}
\affiliation{
PG Department of Physics, 
Rashtrasant Tukdoji Maharaj Nagpur University, 
Campus, Nagpur-440 033, India}

\begin{abstract}

We investigate coupled circle maps in presence of feedback and
explore various dynamical phases observed in this system of coupled
high dimensional maps. We observe an interesting transition from
localized chaos to spatiotemporal chaos. We study this  transition as a dynamic phase
 transition. We observe that persistence acts as an excellent quantifier to 
describe this transition. Taking the location of the fixed point of circle map 
(which does not change with feedback) as a reference point, we compute number of
 sites which have been greater than (less than) the fixed point till time $t$. 
Though local dynamics is high-dimensional in this case, this definition of 
persistence which tracks a single variable is an excellent quantifier for 
this transition. In most cases, we also obtain a well defined persistence 
exponent at the critical point and observe  conventional scaling  as seen 
in second order phase transitions. This indicates that persistence could work as good order
parameter for transitions from fully or partially arrested phase. We also give an 
explanation of gaps in eigenvalue spectrum of the Jacobian of localized state.

\end{abstract}
\pacs{05.45.Ra, 05.70.Fh}

\maketitle

\begin{quotation}
Coupled map lattices have been investigated extensively in past few
decades. The dynamics can show very novel features when local
dynamics is higher dimensional. Using feedback to make the local
dynamics higher dimensional makes it possible to systematically
increase dimensionality of local map without changing several other
features such as location of fixed point. Investigation of phase
diagram of such a system shows possibility of existence of partially
arrested state in which certain sites are near the fixed point while
others are not. We study the nonequilibrium phase transition from such
a state to a state of spatiotemporal chaos and propose that
persistence works as an excellent order parameter for such transitions.
\end{quotation}

\section{\label{sec:level1}Introduction}

Pattern formation in spatially extended systems has been an object of extensive 
study in past few decades.  The reasons for the interest in pattern formation are 
not far to seek. Pattern formation happens in several natural systems and 
scientists are interested in understanding it. Examples could be flame 
propagation or occurrence of spirals in a chemical reactions or patterns on the 
skins of animals modeled by reaction-diffusion processes. Several systems like
biological patterns \cite{biology}, charge density waves or Josephson Junction Arrays
\cite{JJA,bak,Joseph}, lasers \cite{rroy} have been studied extensively from the viewpoint
 of dynamics and pattern formation. The practical importance of understanding 
above systems can not be overemphasized. Partial differential equations, coupled
 ordinary differential equations, coupled oscillator arrays and coupled maps 
have all been used to model different physical and biological systems and have 
uncovered interesting modifications of equilibrium, bifurcations and stability
properties of collective spatiotemporal states. These systems have been studied
 extensively\cite{cross,footnote1}. However, while we have arrived at a good understanding of low 
dimensional chaos in past two decades, and understood routes to chaos in several
seemingly disparate systems in an unified manner, the same is not true for 
spatiotemporal systems.

As one would expect, there is practical relevance to these models.
Certain simplified models of spatiotemporal dynamics have been
studied extensively in recent past. We would like to make a special
mention of coupled map lattices which have been studied extensively
in the hope that the understanding we have developed in
low-dimensional systems could be useful in understanding
spatiotemporal dynamics. Majority of the studies are in coupled
one-dimensional maps. At times, they have been successful in
modeling certain patterns observed in
spatiotemporal systems. For example, the phenomena
 of spatiotemporal intermittency in Rayleigh-Benard convection has been 
modeled by coupled map lattices \cite{chate,bigazzi}. They have been
used to model spiral waves in B-Z reaction \cite{barkley} and even in 
crystal growth \cite{reynolds}.	

 We feel that these studies could be helped considerably by {\em {quantitatively}}
 identifying different phases in the system and attempting to understand 
transitions between those. A lot of attention has been devoted to transition 
to synchronization in these systems which is easy to identify and
analyze. However, other spatiotemporal patterns are far less analyzed
and understood.

It is worth noting that in several spatiotemporal systems the field 
is high dimensional. For example, for a chemical reaction, we would require
information about concentrations of all reactants and products at all
points to be able to simulate the system. There are several phenomena
which are possible only when dynamics is higher dimensional. On the other hand
most of the existing studies in coupled map lattices are about coupled
one-dimensional maps and very few involving high dimensional maps, for example 
coupled Chialvo's map \cite{jampa}, coupled Henon map \cite{politi_chaos92}, Arnold's 
cat map \cite{bonetto_CUP}, modified Baker's map for Ising model \cite{sakaguchi_pre99}, 
etc. Here we try to study coupled higher dimensional maps. We try to systematically 
change the dimensionality of the map and study coupled maps with feedback where 
the feedback time could be varied.  
\begin{figure}
\includegraphics[width=80mm,height=60mm]{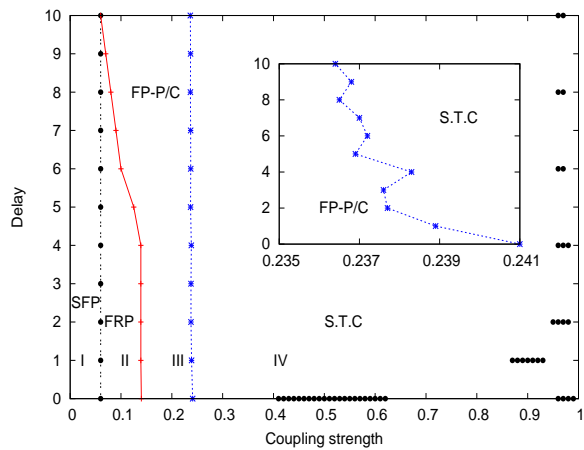}
\caption{\label{fig1} We plot the phase diagram of coupled circle map
with delay in two parameter space of coupling strength $\epsilon$ and
delay time $\tau$. The abbreviation SFP, FRP, FP-P/C and STC denote 
Synchronous fixed point, Frozen random pattern, Fixed point-Periodic/Chaotic 
cycles and Spatiotemporal chaos respectively. In the inset, we magnify the phase 
boundary between FP-P/C regime and STC regime. We can have only integer values of 
feedback time and the lines in phase diagram are guide to the eye. The black solid 
circles indicate points in phase space where spatiotemporal fixed point phase is seen.     
 }
\end{figure}
In coupled map lattice (CML) \cite{kaneko} time evolution
is conflict between competing tendencies: the diffusive coupling between units
tends to make the system homogeneous in space, while the chaotic map produces
temporal inhomogeneity, due to sensitive dependence on initial condition. Such 
systems display a rich phase diagram and origin and stability of various phases 
in these systems are of theoretical and practical interest. The transitions 
between various phases in these systems could be viewed as dynamic phase 
transitions. Though these are clearly non-equilibrium systems, different 
techniques used in the study of equilibrium phase transitions have been applied 
to explore the variety of phenomenologically distinct behaviors. 
Such non-equilibrium systems range from growths of surfaces\cite{stanley} to traffic jams 
in vehicular traffic and granular flows\cite{wolf, haye}. 
For such analysis, there is a need to define an order parameter that will 
characterize the transition. However, not many dynamic phase transitions have 
been studied from this viewpoint \cite{gupte72,gupte74,rahulpandit,gade,ashwini}. 
Most of the studies are devoted to transition in models with so-called absorbing states. 
In the context of synchronization, there have been extensive studies to ascertain 
whether or not this transition is in the universality class of directed 
percolation \cite{ginelli}. Transition to synchronization, locking in of the dynamics 
of several systems to a collective oscillation, is an important but minuscule part 
of overall  spatially extended systems. Several non-synchronization transitions have been 
observed such as spiral state to spatiotemporal chaos \cite{rahulpandit}, 
traveling wave to spatiotemporal chaos \cite{gade}. The spatiotemporal dynamics
 is far richer and other transitions deserve attention. 

 As suggested by Pyragas, feedback in the dynamical systems can play a central role 
in controlling chaos by stabilizing UPO's embedded in chaotic attractor\cite{pyragas}.
 The scheme has been efficiently implemented in wide range of experimental 
applications. Feedback leading  to synchronization is a well established
 fact supported by several researchers\cite{handbook}. Though feedback has been 
an important aspect of studies in control and synchronization, not much has been 
said about its effect on different dynamical phases. It is then important to find 
the effects of feedback in terms of alterations in system dynamics that may lead
 to origin of different dynamic phases. 

 In this work, we investigate coupled maps with feedback. As mentioned before, 
coupled higher dimensional maps have not been studied adequately. Using feedback, 
we can systematically vary dimensionality of local map using feedback and study
 its effect on overall dynamics. We note that the feedback as defined in our model 
does not change the location of fixed point of system without feedback. We have 
used it extensively as a reference point. This allows us to study spatially extended
 higher dimensional systems with known properties. We investigate a model of 
coupled circle maps with feedback. The strength of feedback and the delay time are 
the parameters in the system. We will present a general phase diagram of the system. 
In particular,  we observe a novel transition from {\em{localized chaos to 
spatiotemporal chaos}}. By localized chaos, we mean that the laminar and 
turbulent phases coexist in the same space domain for weaker couplings. 
The reason to call it localized chaos is that one can see there are well defined
 boundaries of turbulent sites which do not mix with laminar ones.
 A similar pattern has been named spatial intermittency (SI) in previous studies and 
some of its properties such as eigenvalue spectrum of Jacobian  are found to be 
related to its dynamical properties\cite{gupte72,gupte74}. Interestingly, we find 
that persistence acts in effective way to characterize and quantify certain
dynamical phases and transition between them in this system. Thus, persistence can be taken as 
excellent order parameter for certain phase transitions. 

 The plan of paper is as follows. We introduce the model in the section II. 
In Section III, we make a detailed analysis of one of the phases, namely localized chaos
and calculate quantities such as turbulent length distribution. In section IV, 
we present our investigations in transition between localized chaos regime and 
spatiotemporal chaos and we present evidence that persistence acts as an excellent 
order parameter to characterize the transition. In section V, we demonstrate  that 
some characteristic features in eigenvalue distributions of Jacobian  can distinguish between 
different dynamical phases. 

\section{Model}
We define  model of coupled circle maps with feedback 
as follows. We assign a continuous variable $x_i(t)$ at each site $i$ at time $t$
where $1\leq \; i \leq N$. $N$ is the number of lattice points. The evolution of
 $x_i(t)$ is defined by,
\begin{eqnarray}
x_i(t+1)=(1-\epsilon -\delta)f(x_i(t)) \;+\;{\frac{\epsilon}{2}} \;[f(x_{i-1}(t)) \nonumber\\ + f(x_{i+1}(t))] 
+ \delta f(x_i(t-\tau))
\label{eq1}
\end{eqnarray}

The parameter $\epsilon$ is the coupling strength and the
local dynamical update function $f(x)$ is the sine-circle map,
\begin{equation}
f(x)=x+\omega-{\frac{k}{2\pi}}\;\sin(2\pi x)
\end{equation}
Where, $k$ is the strength of nonlinearity and $\omega$ is the winding number. 
The  parameter $\delta$ is strength of feedback and $\tau$ is delay time.
 If we increase $\tau$, we can increase the dimension of map. In our simulations,
we confine values of parameters to $\delta = 0.1$, $k = 1$ and $\omega = 0.068$. 
The coupling topology considered here is diffusive i.e nearest - neighbor connections.
 For coupled circle map, we confine the dynamics in the interval $[0,1]$ using
 the following rule. If int$[x_i(t)] = m$, $x_i(t) = x_i(t)-m$ if $x_i(t) > 0$ 
and $x_i(t) = x_i(t)-m+1$ if $x_i(t)< 0$. We study the system with random initial
 conditions. The system is updated synchronously with periodic boundary condition.
 The feedback does not change the fixed point
 solution of local map  $f(x)$ without feedback. The fixed point solution for 
the local map $f(x)$ is given by,
$x^*={\frac {1}{2\pi}} \sin^{-1}({\frac{2\pi\omega}{k}})$.
 The parameter 
values considered here yield the value of fixed point as $x^* = 0.07026$.
This point acts as an excellent reference point in visualizing the 
dynamical behavior of system and also in symbolic dynamics of the system.
 We can define the site dynamics to be laminar if $\vert x_i(t)-x^* \vert <\eta$ 
for small enough neighborhood $\eta = 0.001$ and to be turbulent otherwise.

\section{Spatiotemporal Dynamics}

\begin{figure}
\includegraphics[width=60mm,height=60mm]{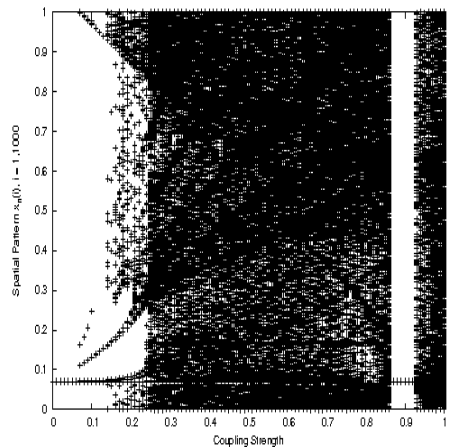}
\caption{\label{fig2} Bifurcation diagram of Spatial pattern at an instant of time 
(after transience), with respect to coupling strength $\epsilon$ for delay time 
$\tau = 1$.}
\end{figure}

We examine the phenomenology of delay coupled circle map with feedback under
variation of coupling strengths and delay time. Various spatiotemporal patterns
are clearly evident from phase diagram Fig. \ref{fig1} and the bifurcation diagram 
shown in Fig. \ref{fig2}. Note that on y axis of Fig. \ref{fig1}, the time is 
discrete. The delay time $\tau$ can only be some integer. The phase boundary of various 
regions can be seen. 

Increasing value of $\tau$ is equivalent to increasing the dimensionality of local dynamics.
There are clearly four distinct behaviors shown by the system as seen from phase 
diagram Fig. \ref{fig1}. The first region in phase diagram labeled as SFP 
(Synchronous fixed point) where, for smaller coupling 
strengths $\epsilon \leq 0.06$ the entire system goes to local
fixed point $x^*$. This  is an absorbing state and system settles in
 this state very quickly. We have checked that this state exists at least  for $\tau \leq 10$.
For higher coupling strengths, The system goes to spatiotemporal chaos as shown 
in phase diagram marked as STC. These different behaviors can also be visualized using
the bifurcation diagram Fig. \ref{fig2} obtained for different coupling 
strengths for $\tau = 1$. This bifurcation diagram shows the state of all sites 
at a given instant of time. For very high coupling strengths, the system shows
some stability islands for some values of $\tau$ though they constitute a small
part in phase diagram, we find them worth a mention. 

 In the intermediate coupling strengths, the system shows a rather interesting behavior. 
We can see that some of the sites approach local fixed point and some of the sites 
take on different values which do not change in time. This behavior is shown in phase diagram
 marked as FRP which stand for frozen random pattern \cite{frozen}. This region is characterized by a spatial
sequence of different domains whose size and periodicity may greatly vary. Upon further increase of 
coupling strengths, one enters in a region where we can see coexistence of different
 dynamical regimes. Where some of the sites indeed go to fixed point, some of the 
sites show periodic/chaotic behavior. This is an interesting state where one can
 see that different dynamical phases do not mix with each other. 
The regions marked as FRP and FP-P/C in the phase diagram collectively represent the 
dynamics which we refer to as {\em Localized chaos}. Chaos is localized in 
certain dynamical regimes of turbulence and does not interfere with evolution of
 laminar states. In localized chaos regime, chaotic domains and periodic domains
 are fixed in space. This is evident from spatial distribution pattern. The 
perturbations do not propagate in space. Essentially, the dynamics is non-spreading 
and non-infective in nature, i.e. the spatially intermittent solutions have zero
 velocity components in spatial direction, and modes traveling along the 
lattice are absent. This can also be considered as a case of spatial intermittency \cite{gupte72}. 
The important result is that spatial intermittency is seen over a large parameter regime in presence of feedback. 
For each feedback time, we have a well defined region where we get the transition. These spatiotemporal patterns are 
clearly evident from space-time density plots 
below. Each figure represent a particular region in phase diagram.

\begin{figure*}
\centering
 \begin{tabular}{l l l}
 \includegraphics[width=60mm,height=60mm]{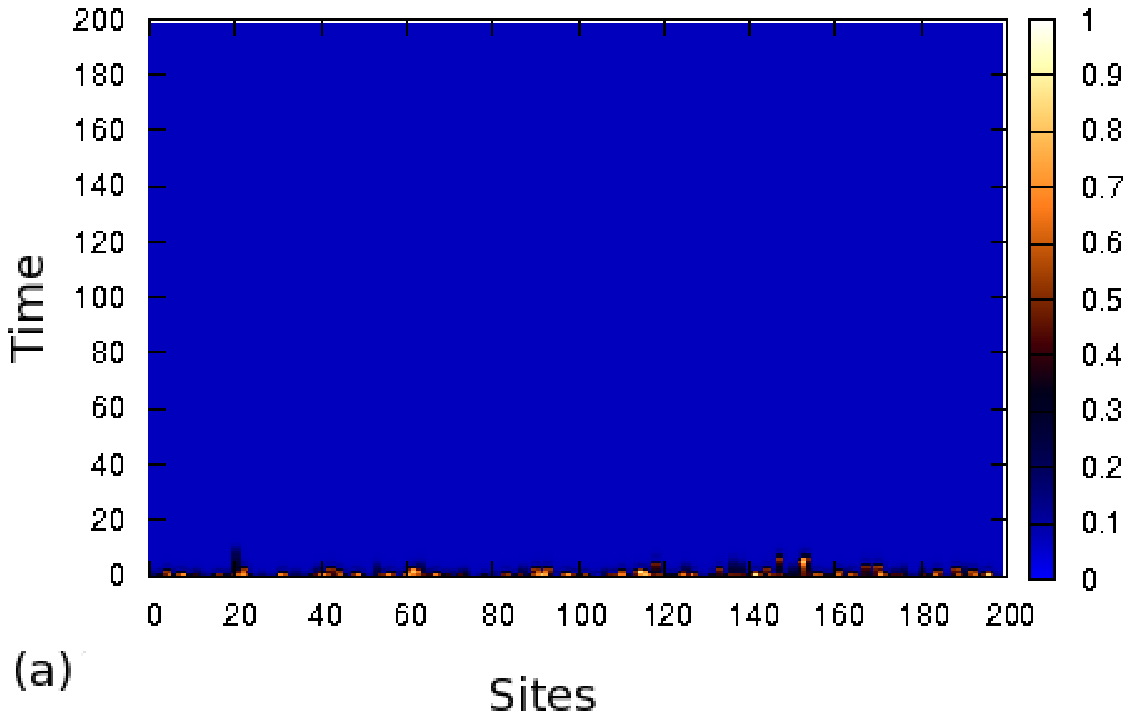} & \includegraphics[width=60mm,height=60mm]{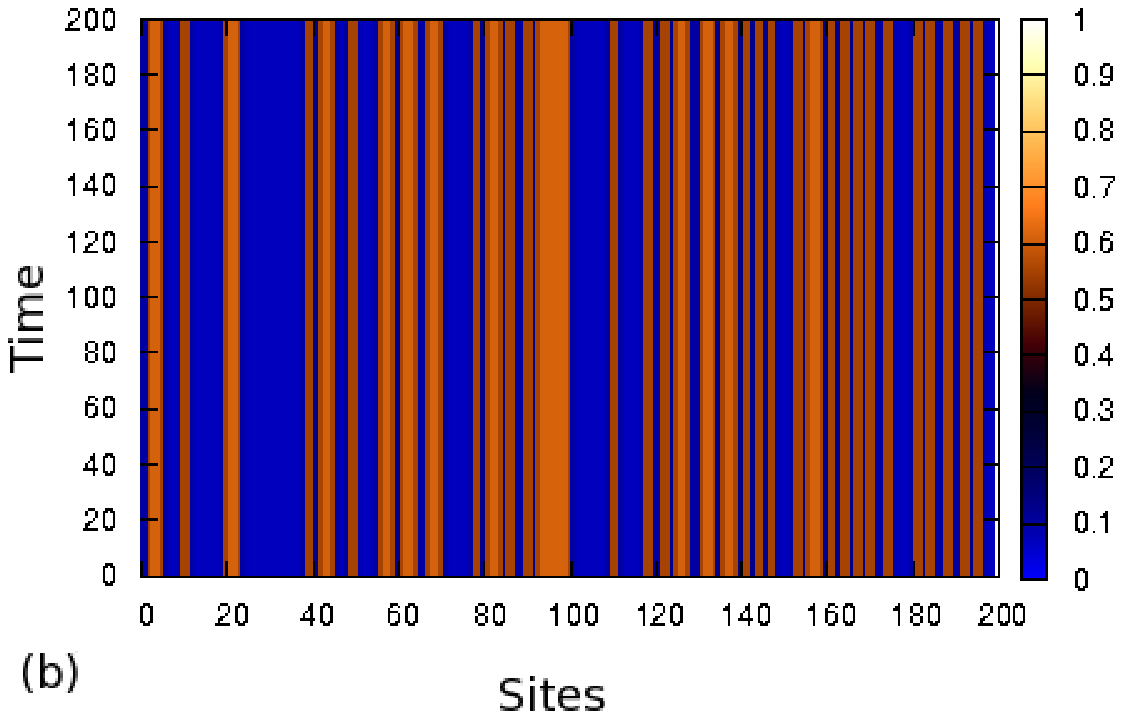} &  \includegraphics[width=60mm,height=60mm]{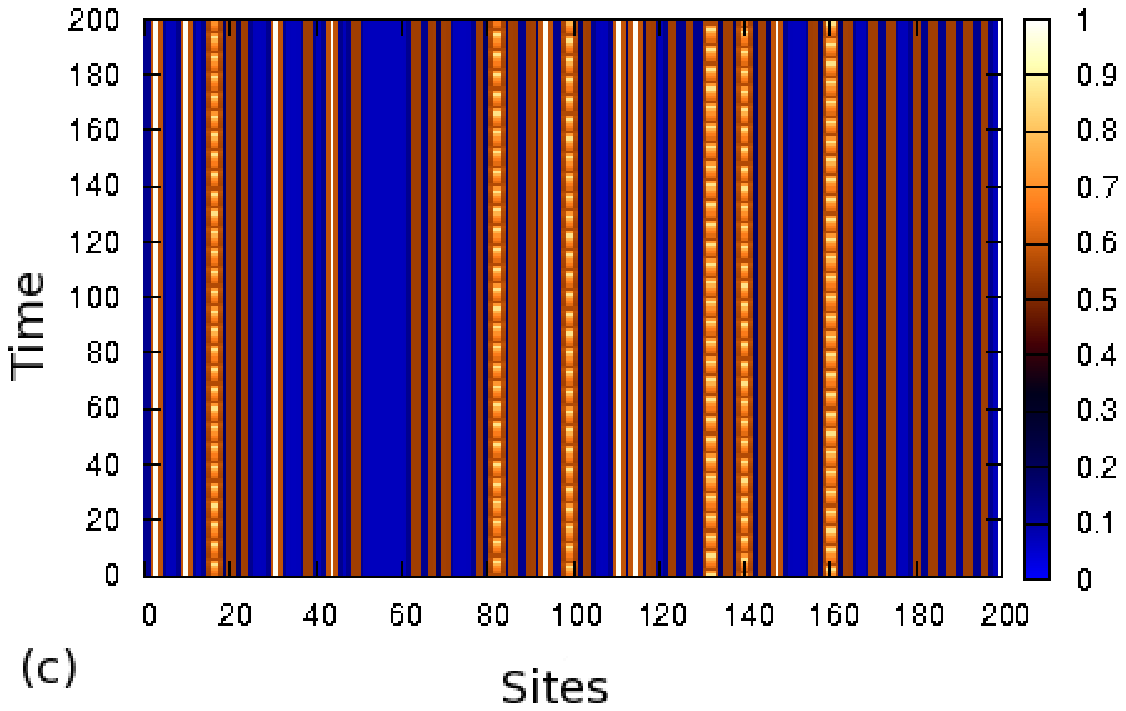} \\
 \includegraphics[width=60mm,height=60mm]{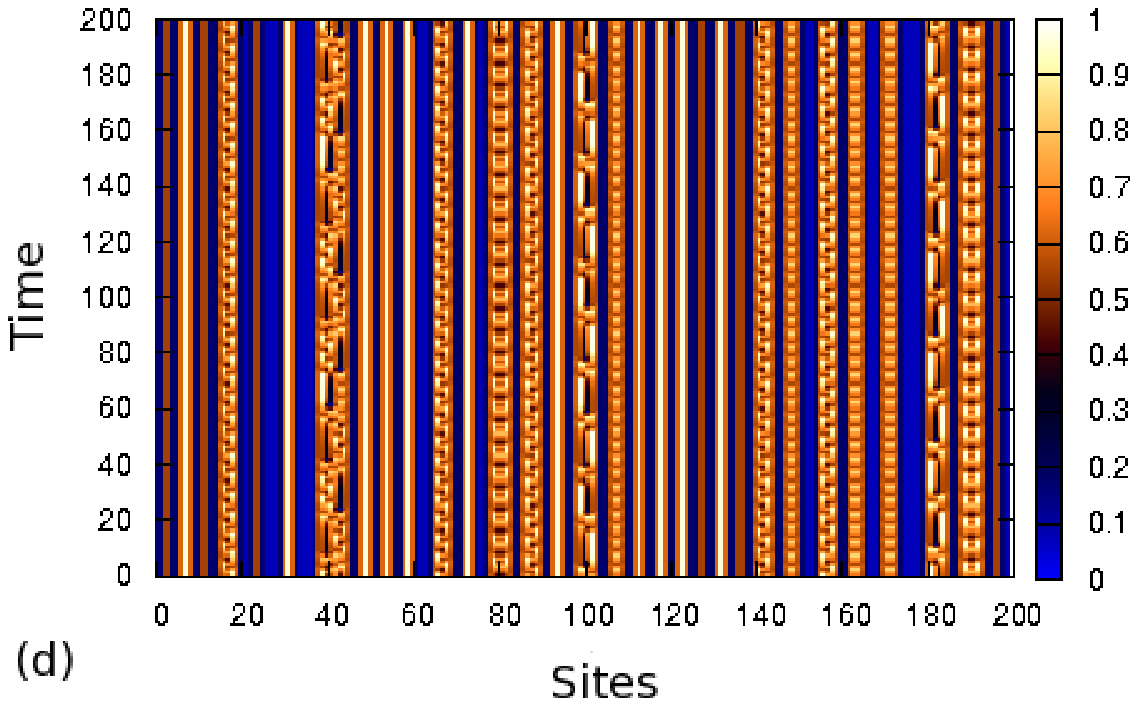} & \includegraphics[width=60mm,height=60mm]{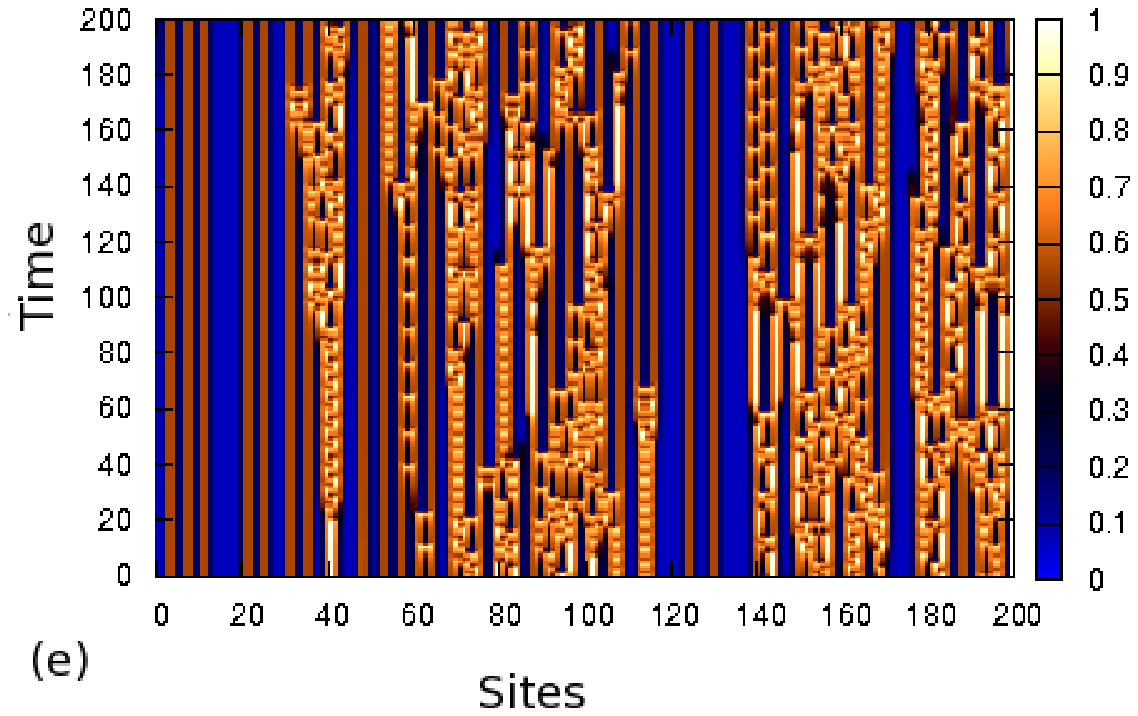} & \includegraphics[width=60mm,height=60mm]{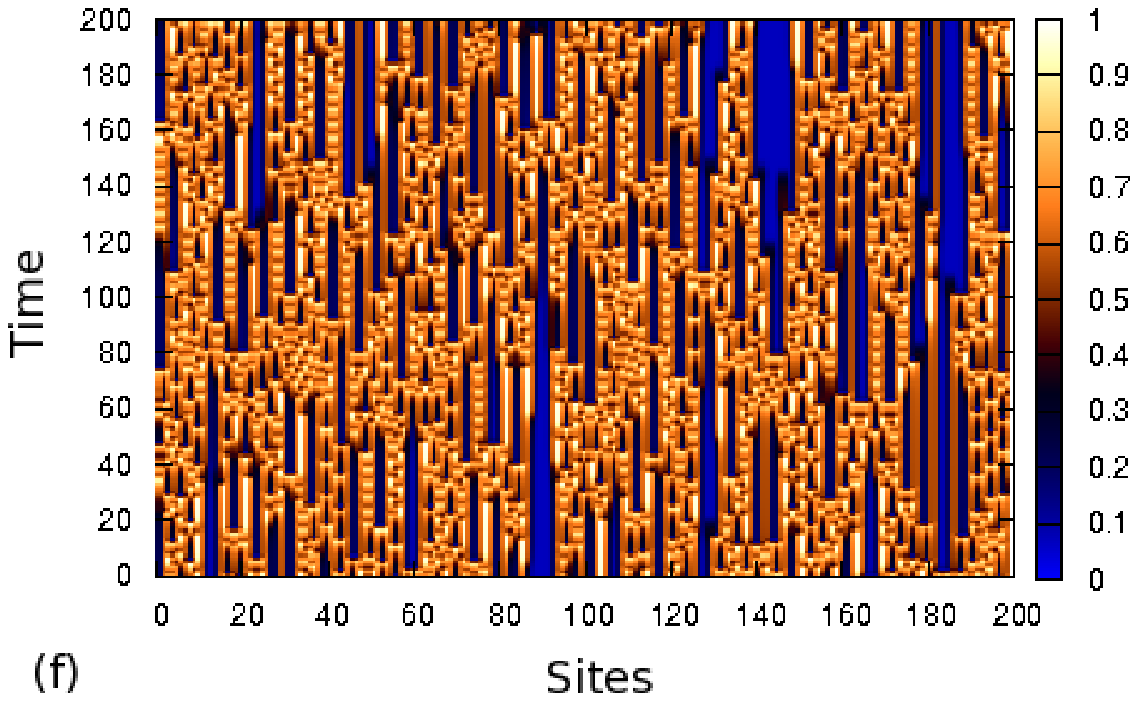} \\
 \end{tabular}
\caption{ \label{fig3}(Color online)Space time density plot of coupled circle map lattice with 
feedback defined through eqs. (1)-(2), in a system of size $L = 200$, for 
parameters $k=1$, $\delta = 0.1$, $\omega=0.068$ for $\tau = 1$ 
and (a)$\epsilon = 0.05$, (b) $\epsilon= 0.1$, (c)$ \epsilon = 0.16$, 
(d) $\epsilon =0.23$,(e) $\epsilon = 0.24097$, and (f) $\epsilon = 0.245$}
\end{figure*}
Fig. \ref{fig3}  shows space-time density plots of $x_{i,t}$ for six choices of parameter values. 
The site index is plotted along the x-axis and the time evolves along the y-axis.
 As mentioned previously, laminar regions are blue (dark grey) in this representation. 
The colorbar shows different value of $x(i)$ according to given scale. Fig. \ref{fig3}(a) 
gives the space-time density plot for parameter values where the system goes to spatiotemporal 
fixed point. Fig. \ref{fig3}(b) shows the plot for non infective and non spreading behavior
called as frozen random pattern. Fig. \ref{fig3}(c) and (d) shows coexistence of different dynamical 
regimes for which the dynamics essentially remains `arrested'. Fig.\ref{fig3}(e) shows density plot for 
$\epsilon = 0.24097$ which is critical point for the transition, we can see the 
spreading behavior just emerging in the system. Fig.\ref{fig3}(f) depicts fully 
grown spatiotemporal chaos and here we can see that the state variable encompasses 
much larger part of available phase space. 

\begin{figure}
\includegraphics[width=80mm,height=80mm]{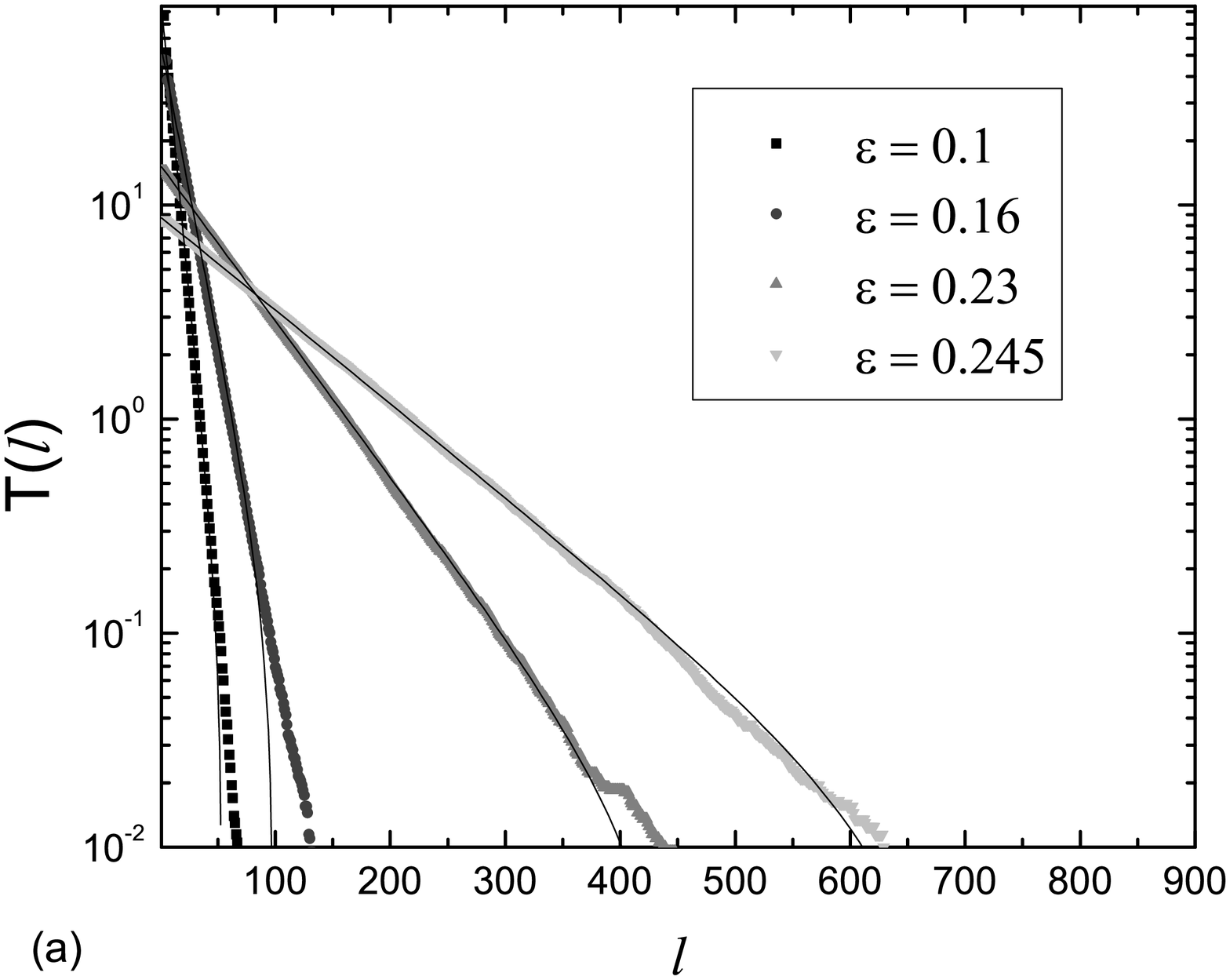}
\includegraphics[width=80mm,height=80mm]{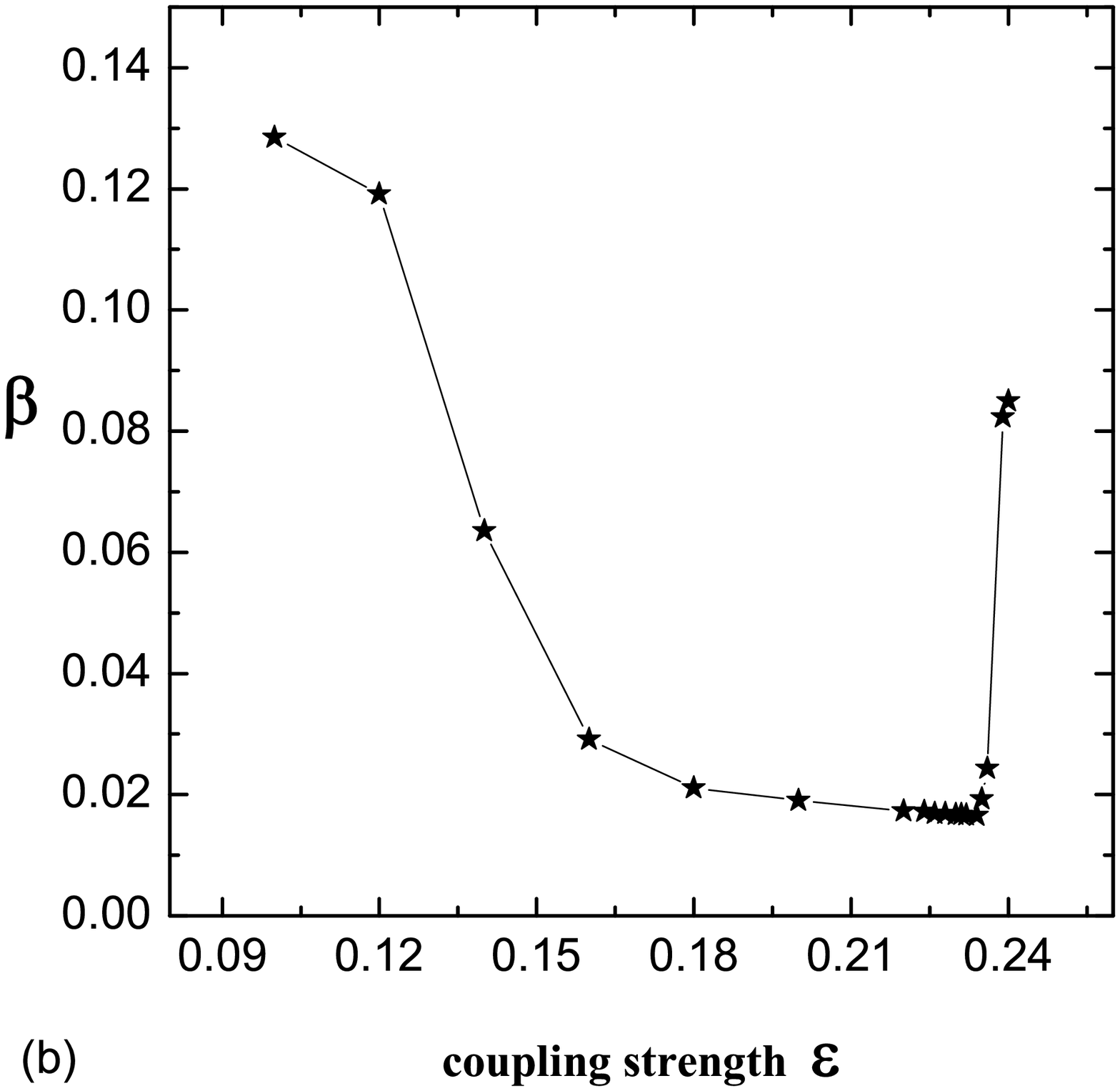}
\caption{\label{fig4}(a) The distribution of turbulent phase lengths for
coupling values $\epsilon = 0.14,0.16,0.23,0.245$ for $\tau = 1$.
The fit to this distribution is exponential with exponent $\beta$. The ordinate axis
is presented in logarithmic scale.
(b) This plot shows exponent $\beta$ vs coupling strength $\epsilon$ for $\tau = 1$.  }
\end{figure}

The number of turbulent sites increase as coupling strength increases. Note here that, all those 
sites which are not laminar are considered turbulent. We calculate distribution of turbulent phase length.
 It is observed that turbulent length distribution is governed by the exponential law,
\begin{equation}
T(l)= A \exp(-\beta l)
\end{equation}
where, exponent $\beta$ is the rate of  decay and $l$ is the 
the cluster size of turbulent sites.
Fig. \ref{fig4}(a) shows Cumulative distribution of turbulent lengths for coupling 
parameter values $\epsilon = 0.1,0.16,0.23,0.245$. We can see that the distribution
 for smaller coupling strengths falls of quickly as opposed to higher value of 
coupling strength where the exponential decay is slower indicating large number 
of turbulent phases. Fig. \ref{fig4}(b) shows how the exponent $\beta$ varies with 
respect to $\epsilon$. There is a sharp increase in value of beta near 
$\epsilon \sim 0.24$ .
  
\section{Persistence as an Order Parameter}

As it can be seen the system displays a dynamical phase transition i.e. from 
localized chaos regime to spatiotemporal chaos. It is important to find  a good 
quantifier to quantify this transition. This is a partially arrested state and 
in partially or fully arrested dynamical states, we believe that persistence acts 
as a good order parameter. There have been couple of previous studied in 
persistence in coupled map lattices by Menon {\it {et al.}} and by Gade {\it {et al.}}.
Menon {\it {et al.}} found that it was a good indicator for a transition from fully
synchronous state to spatiotemporal chaos while Gade {\it {et al.}} found that its 
dynamical behavior is able to characterize even traveling wave state which can be seen 
as a partially arrested state in moving frame of reference\cite{menon, gade}. It was 
observed by us in preliminary version of these studies \cite{abhijeet_dae} that 
persistence acts as an good quantifier. Recent results by Mahajan and Gade show 
that local persistence can also characterize the transition from clustered state 
to spatiotemporal chaos in small world networks\cite{ashwini}. This definition of persistence concentrates 
on a single variable and one would wonders if it is adequate for the high dimensional local
dynamics we are studying in this case. However, we find that persistence acts as an 
excellent order parameter for distinguishing between partially and fully arrested 
states in this system. This demonstrates a generic utility of persistence as 
order parameter even in transitions from partially arrested phase. Thus, it is 
possible that persistence is a good characteristic for  systems in which dynamics is 
partially or fully localized. Conventionally, persistence in the context of 
stochastic processes has been defined as the probability $P(t)$ that a stochastically 
fluctuating variable with a zero mean has not crossed a threshold value up to 
time $t$ \cite{satya}. For a spin system with discrete states, such as Ising or 
Potts spins, it is defined in terms of the probability that a given spin has not flipped
out of its initial state up to time $t$.  The power-law tail of $P(t)$,
{\it i.e.\/}
\begin{equation}
P(t) \sim 1/t^\theta,
\end{equation}
defines the persistence exponent $\theta$. The exponent $\theta$ is 
dimension-dependent, non-trivial exponent. For spatially-extended systems, the 
time evolution of the stochastic field at one lattice site is coupled to its 
neighbors. This makes the effective single-site evolution essentially 
non-markovian and renders the nontriviality to the exponent $\theta$. This nature 
of local persistence probability relates it to infinite point correlation function 
in time direction \cite{koduvely}. 
\begin{figure}
\includegraphics[width=80mm,height=80mm]{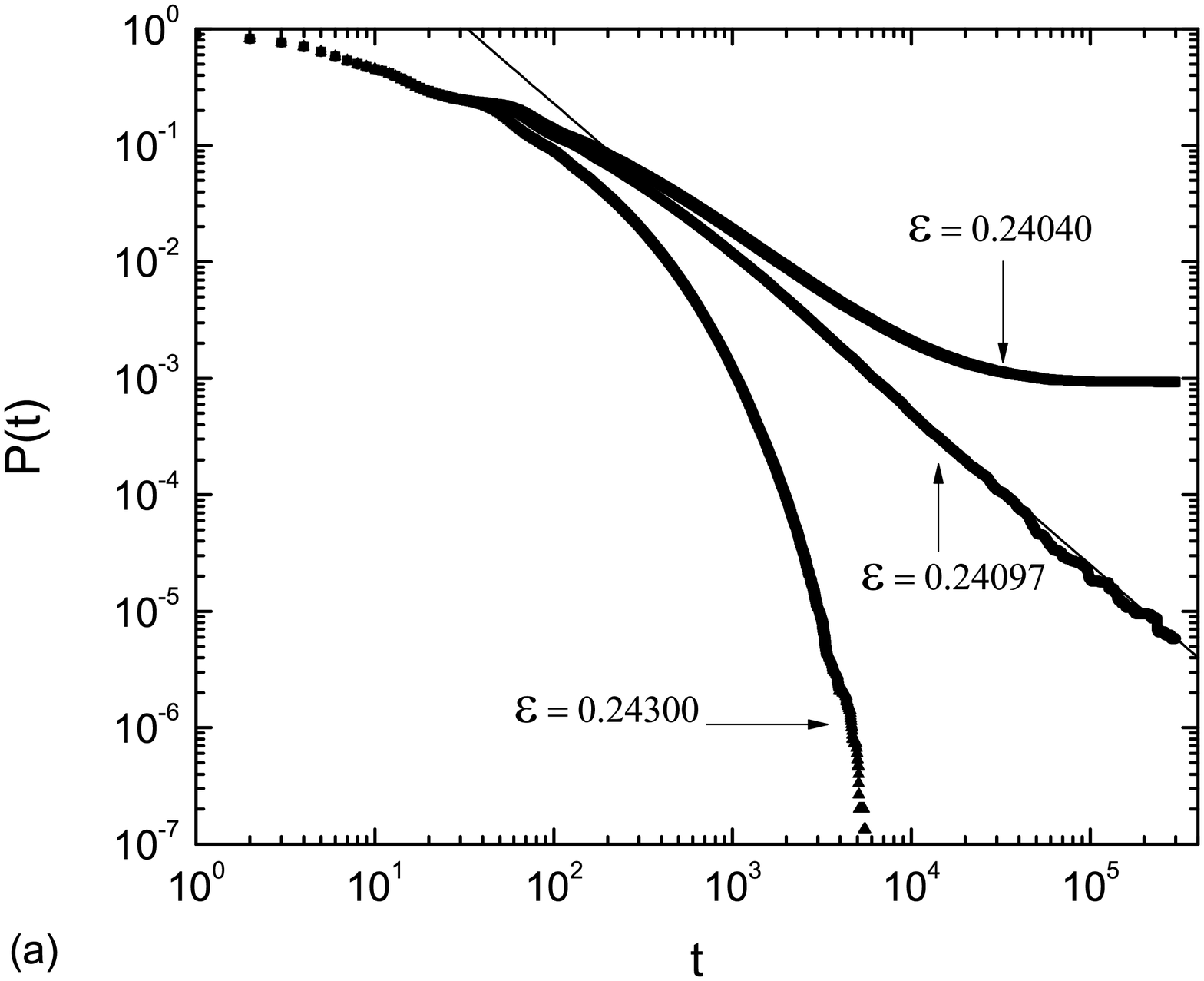}
\includegraphics[width=80mm,height=80mm]{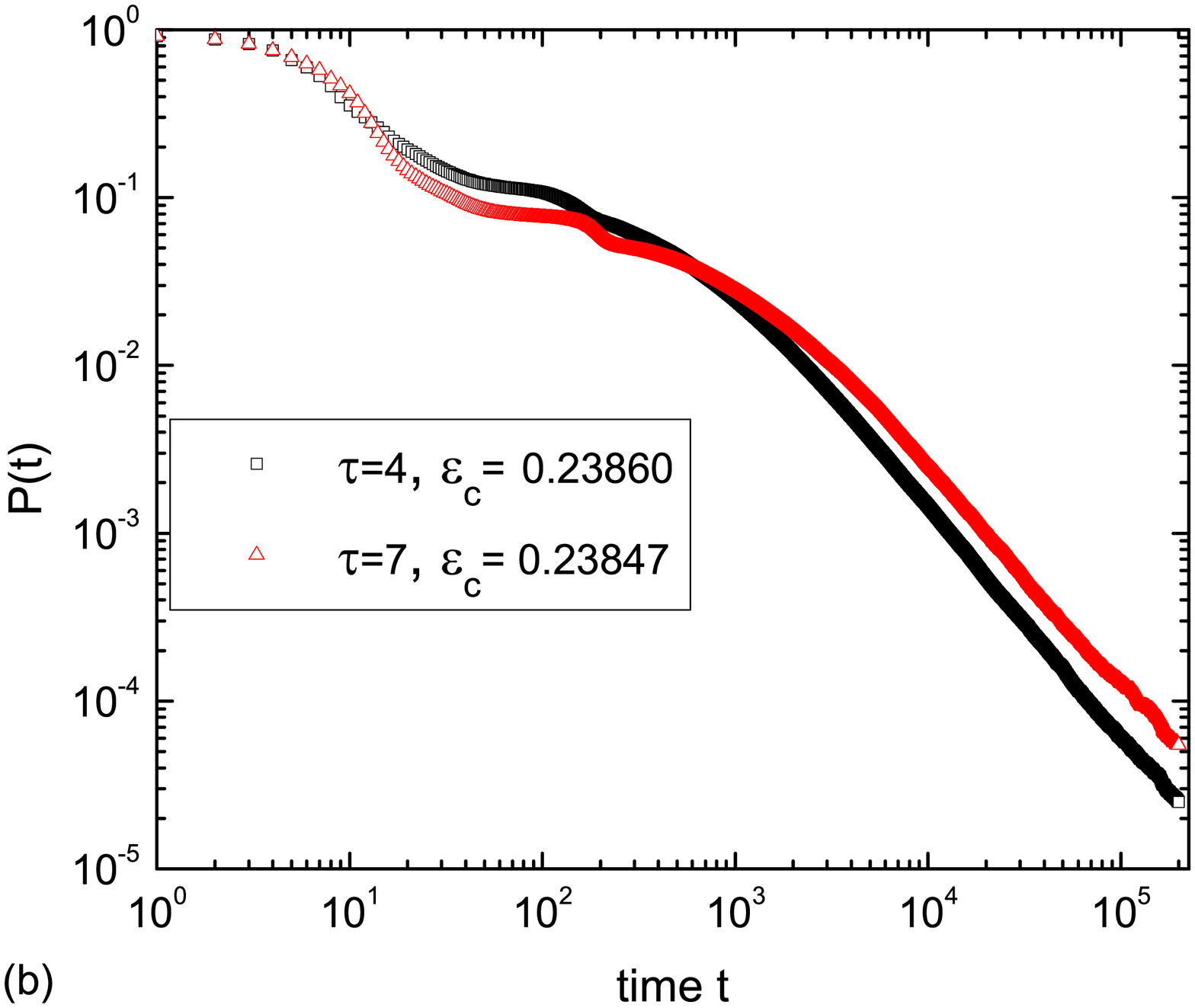}
\caption{\label{fig5} (a) Persistence probability $P(t)$ is plotted as a function of 
time $t$ on a logarithmic scale for $k= 1$, $\omega=0.068$ and $\tau = 1$. The lattice size is 5000.
 The curves shown are for
 $\epsilon= 0.24040$ (upper), $\epsilon= 0.24097$ (middle), and $\epsilon=0.24300$ (lower).
The line with slop 1.31 is plotted for reference.
(b) (Color online) Log P(t) vs Log t for $\tau = 4$, $\tau = 7$.}
\end{figure}

 A natural generalization of the idea to CML as suggested by Menon {\it {et al.}} 
\cite{menon}, defines local persistence in terms of the probability that a 
local state variable $x_{(i,t)}$ does not cross fixed point value $x^*$ up to 
time $t$. With this definition, we study persistence in the coupled map lattice
defined through eq (\ref{eq1}), via simulations on systems with various lattice 
sizes. The persistence defined above is referred as local persistence in 
literature. In this work, we study only local persistence and do not study any 
other definition of persistence. In this paper, by persistence we mean local 
persistence only. We start with local $x_i$ distributed randomly over interval $[0,1]$. 
We find the normalized fraction of persistent sites, defined as sites for 
which $x_{i,t}$ has not crossed $x^*$ up to time $t$. Thus, sites for which the sign 
of $(x_{i,t} -x^*)$ has not changed unto time $t$ are persistent at time $t$. This 
persistence probability $P(t)$ is averaged over an ensemble of over 2000 random initial 
conditions. 

 Fig. \ref{fig5}(a) shows plots of P(t) at three different values of coupling 
parameter $\epsilon$. It can be easily noted that for values below critical 
point $P(t)$ saturates. For the value above critical point i.e. in chaotic 
domain, $ P(t)$ decays exponentially. Exactly at the critical point of transition
 between localized chaos to spatiotemporal chaotic regimes, $P(t)$ decays as
 a power law, which defines the persistence exponent  $\theta  = 1.31$. The power law
decay can also be seen for other values of $\tau$ \cite{footnote2} . Fig. \ref{fig5}(b) shows persistence 
distribution for $\tau = 4, 7$ at the critical points $\epsilon_c = 0.2386$ and $\epsilon_c = 0.23847$
respectively.
 The observed power-law behavior in 1+1 dimension suggests the further investigation of 
parameters involved in phenomenological scaling laws. 
 The distance from criticality $\Delta = |\epsilon - \epsilon_c|$ and lateral system 
size $L$ can be used as parameters to study scaling properties of the system.
One would expect a asymptotic scaling law of the form
\begin{equation}
P(t) \simeq  t^{-\theta}F\Big(tL^{-z},t\Delta^{\nu_{||}} \Big)
\end {equation}
where, $F$ is a scaling function and $z = \nu_{||}/\nu_\perp = 1.62$ is the 
dynamical exponent. We consider two cases to demonstrate the 
validity of this scaling form numerically.
\begin{figure}
\includegraphics[width=60mm,height=60mm]{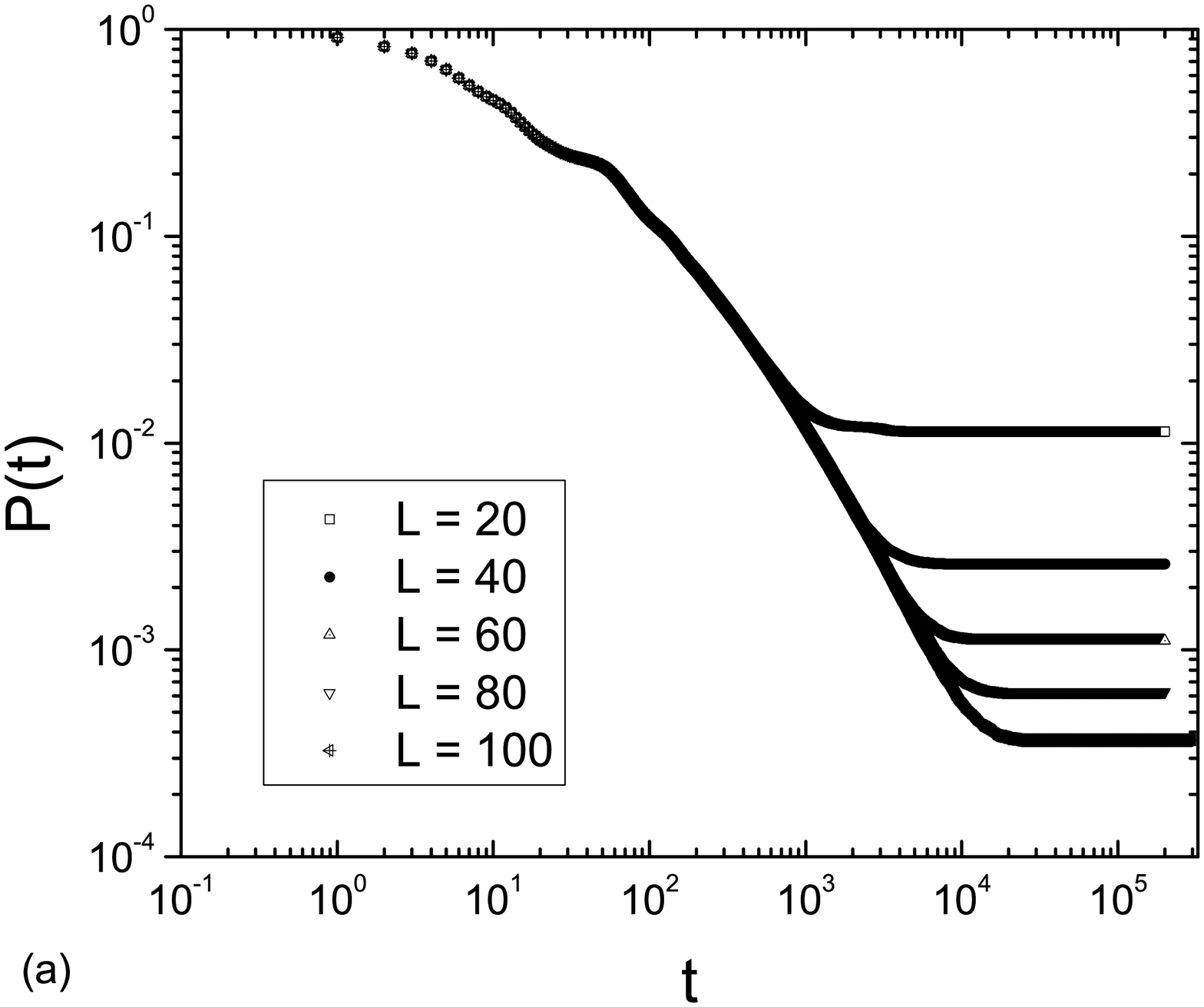}
\includegraphics[width=60mm,height=60mm]{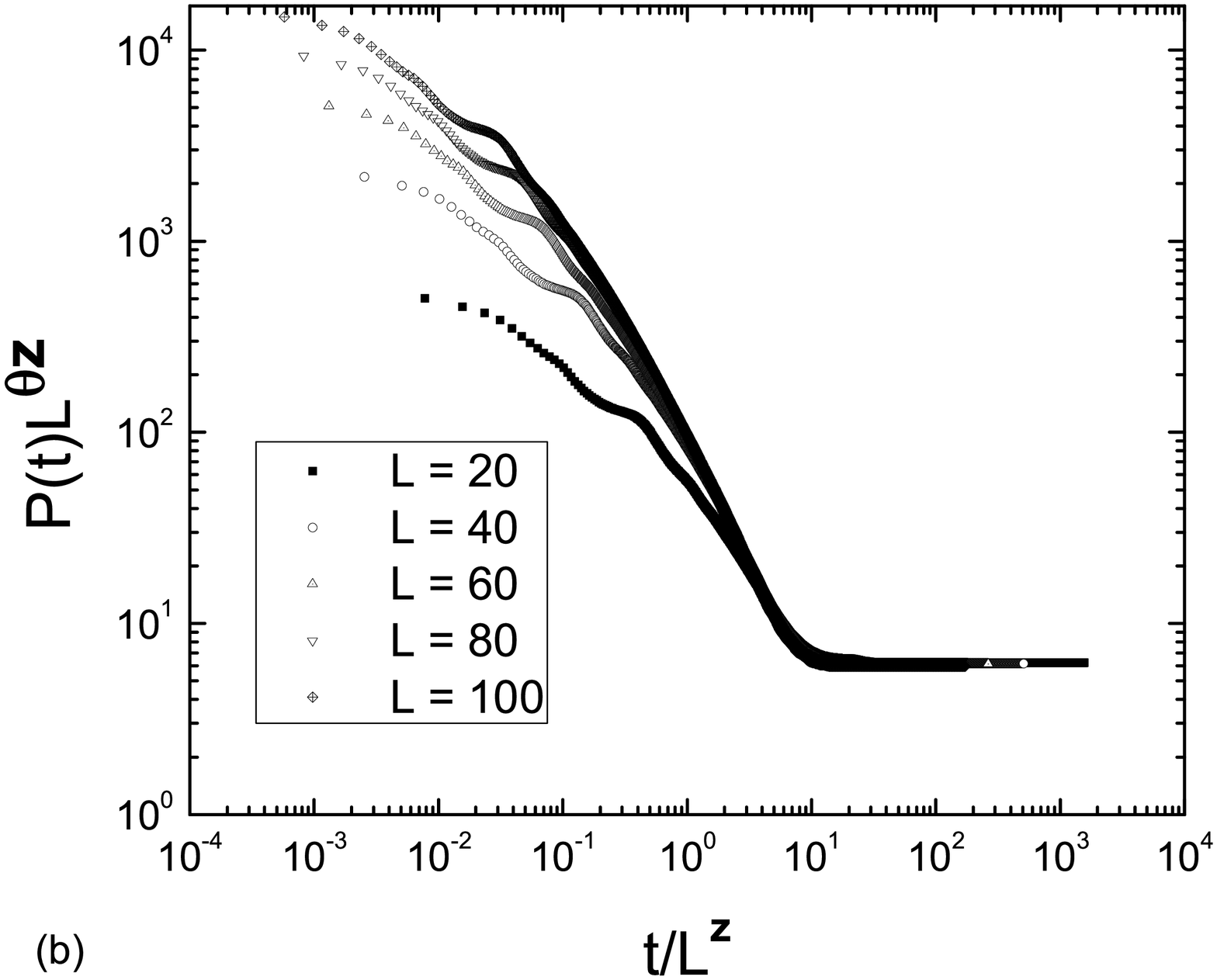}
\caption{\label{fig6}(a) The persistence probability $P(t,L)$ is plotted against
time $t$ for different lattice sizes $L$.
(b)The scaling function$f(x) = P(t,L)L^{z\theta}$ is plotted against 
dimensionless scaling variable $x = tL^{-z}$. The data for different $L$ values 
were found to collapse for longer times for $z = 1.62$.  }
\end{figure}

 As stated by Fuchs et. al.\cite{haye2}, a finite 1+1 dimensional contact process at criticality 
reaches the absorbing state within finite time so that there is a finite probability
 for persistent sites to survive forever. Therefore, persistence probability 
saturates at a constant value. One can easily carry this analogy to non spreading
 regime in coupled evolution of delayed circle maps in localized chaos regime.
When the dynamics is such that no site is laminar, the persistence decays to zero. The scaling function at criticality $\Delta = 0$
 for various system sizes L, i.e.
plotting $P(t)L^{\theta z}$ against $t/L^z$, shows a excellent data collapse
 at longer times for $z = 1.62$ and $\theta = 1.31$ in Fig. \ref{fig6}.

\begin{figure}
\includegraphics[width=60mm,height=60mm]{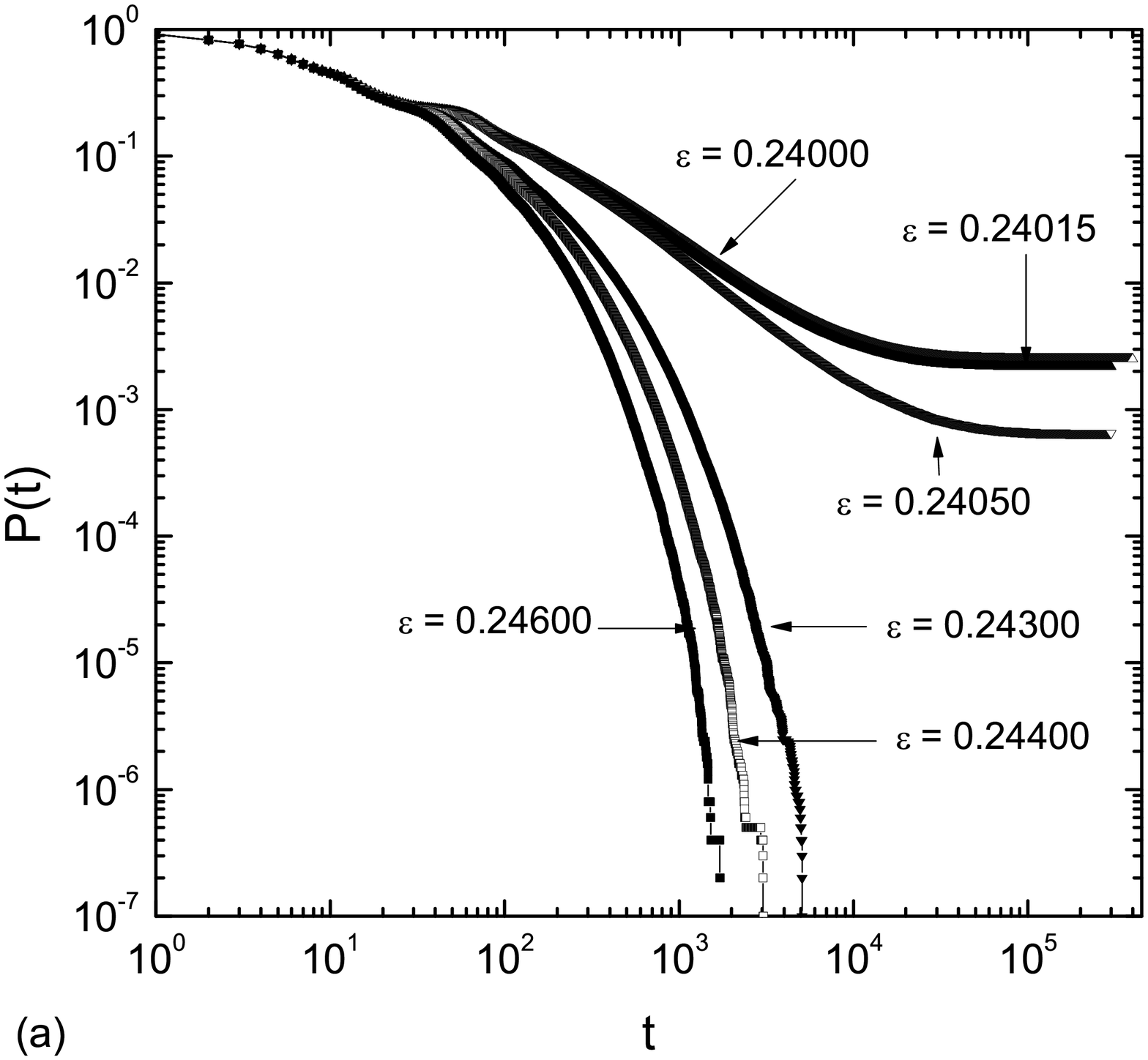}
\includegraphics[width=60mm,height=60mm]{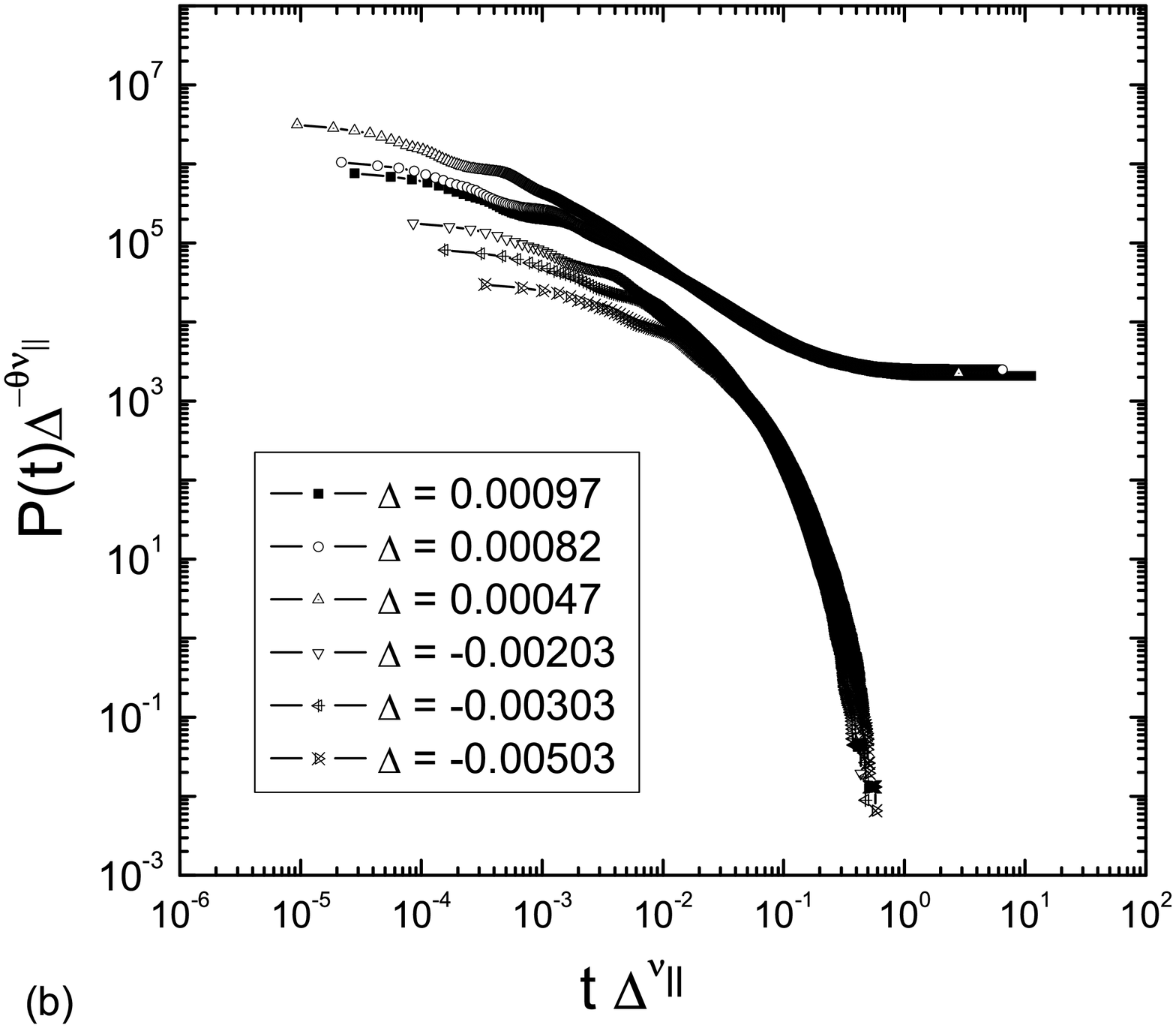}
\caption{\label{fig7}(a) The persistence probability $P(t,L)$ is plotted against
time $t$ for different coupling strengths $\epsilon$.
(b) The scaling function$f(x) = P(t)\Delta^{-\theta \nu_{||}}$
 is plotted against dimensionless 
scaling variable $x = t\Delta^{\nu_{||}}$. A good data collapse is obtained for 
 $\nu_{||} = 1.51$.  }
\end{figure}
\begin{figure}
\includegraphics[width=60mm,height=60mm]{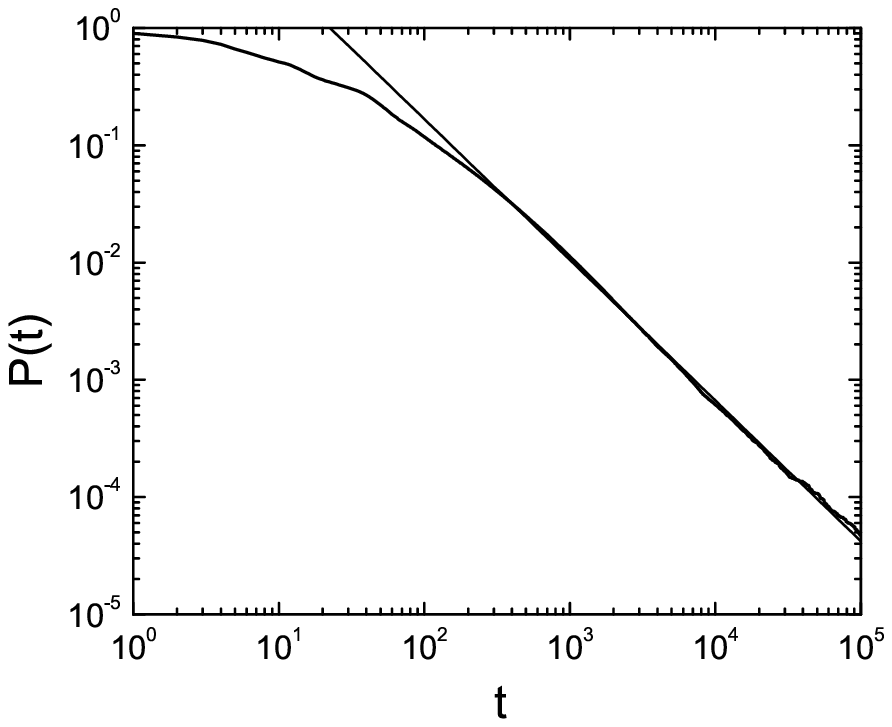}
\caption{\label{fig8} (a) Persistence probability $P(t)$ is plotted as a function of 
time $t$ on a logarithmic scale for $k= 1$, $\omega=0.062$ and $\epsilon = 0.26634$. 
The lattice size is 5000. The persistence exponent $\theta =1.2$.}
\end{figure}
For a continuous phase transition below (above) the critical point, we expect the 
persistence probability to saturate (decay exponentially). According to above 
scaling form, the curves should collapse in both cases when $P(t)\Delta^{-\theta \nu_{||}}$ is plotted against 
$t\Delta^{\nu_{||}}$. This is indeed the case as shown in Fig. \ref{fig7}. The simulations
were carried out for lattice size $L = 5000$. The size is large enough so that finite size 
corrections could be neglected. Also ensemble averaging has been performed over 5000 different
initial conditions to ensure convergence. Persistence acts as an excellent order 
parameter for transition from spreading to nonspreading regimes. This point can further 
be substantiated by similar power law decay seen for $\delta = 0$ case of spatial 
intermittency, as seen by Jabeen and Gupte \cite{gupte74}. Figure  \ref{fig8} shows 
persistence probability behavior for parameter values of coupled circle map without 
feedback where spatial intermittency is seen. The persistence exponent observed is $1.2$. 
We would like to note that persistence exponent obtained in transition from laminar 
state to spatiotemporal chaos (shown to be in DP universality class) is found to 
$1.5$ \cite{menon} while in transition from laminar state to traveling wave state in negatively 
coupled circle maps, the exponent is $1$ \cite{gade}. Thus the exponent is clearly different 
from the exponents observed in previously studied transitions. However as noted previously in 
literature, persistence is known to be least universal exponent among various 
dynamical exponents.

\section{Stability Analysis}
Computing Jacobian matrix of the system is useful from several
viewpoints. Lyapunov exponents are essentially related to eigenvalues
of product of Jacobians over infinite period. For a synchronized or
periodic solutions, one can get considerable simplifications by
analyzing the Jacobian matrix for those states. The localized state
has been previously observed in coupled circle maps by Jabeen and Gupte \cite{gupte74}
and has been  named as spatial intermittency and they observed certain
differences between eigenvalue distributions of one-step Jacobian
matrix. It has been observed that eigenvalue spectrum of the Jacobian of the
system at any time $T$ gives a good indicator of this phase. We show
that the eigenvalues remain real for our systems as well. 
 This spectrum shows distinct features in the phases concerned.

For $\tau=1$, our equations can be put in the form.
\begin{eqnarray*}
x_i(t+1) &=& (1-\epsilon -\delta)f(x_i(t)) \;+\; {\frac{\epsilon}{2}} \;[f(x_{i-1}(t)) \\
& & + f(x_{i+1}(t))] + \delta f(y_i(t))\\ \nonumber
y_i(t+1) &=& x_i(t)
\end{eqnarray*}
The Jacobian $J_t$ at time $t$ is given by,
\begin{displaymath}
\mathbf{J_t} =
\left(\begin{array}{cc}
A_t & B_N \\
I_N & 0_N
\end{array}\right)
\end{displaymath}
where $0_N$ is an $N \times N$  Null matrix with all entries $0$. $I_N$ is 
$N$-dimensional identity matrix. The matrix $B_N$ is defined as $B(i,i) = \delta f'(y_i(t))$
and $B(i,j) = 0$, if $i\neq j$.
The matrix $A_t$ is given by,
 \begin{widetext}
\begin{displaymath}
\mathbf{A_t} =
\left(\begin{array}{ccccc}
\epsilon_s f'(x_1(t)) & \epsilon_n f'(x_2(t)) & 0 & \ldots & \epsilon_n f'(x_N(t)) \\
\epsilon_n f'(x_1(t)) & \epsilon_s f'(x_2(t)) & \epsilon_n f'(x_3(t)) & \ldots & 0 \\
0 & \epsilon_n f'(x_2(t)) & \epsilon_s f'(x_3(t)) & \epsilon_n f'(x_4(t)) & \ldots \\
\vdots & \vdots & \vdots & \vdots & \vdots \\
\epsilon_n f'(x_1(t)) & 0 & \ldots & \epsilon_n f'(x_{N-1}(t)) & \epsilon_s f'(x_N(t))\\
\end{array}\right)
\end{displaymath}
\end{widetext}
 where, $\epsilon_s$ = $1-\epsilon-\delta$, $\epsilon_n = \frac{\epsilon}{2}$ and 
$f'(x_i(t)$ = $1 - K cos(2\pi x_i(t))$. where, $x_i(t)$ is the state variable. 
The diagonalization of $J_t$ gives the $2N$ eigenvalues of the stability matrix. 
The eigenvalues of the stability matrix were calculated for values of state 
variable $x_i(t)$ for all $i$ at one time step on discarding sufficiently long
transients.

\begin{figure*}
\includegraphics[width=70mm,height=70mm]{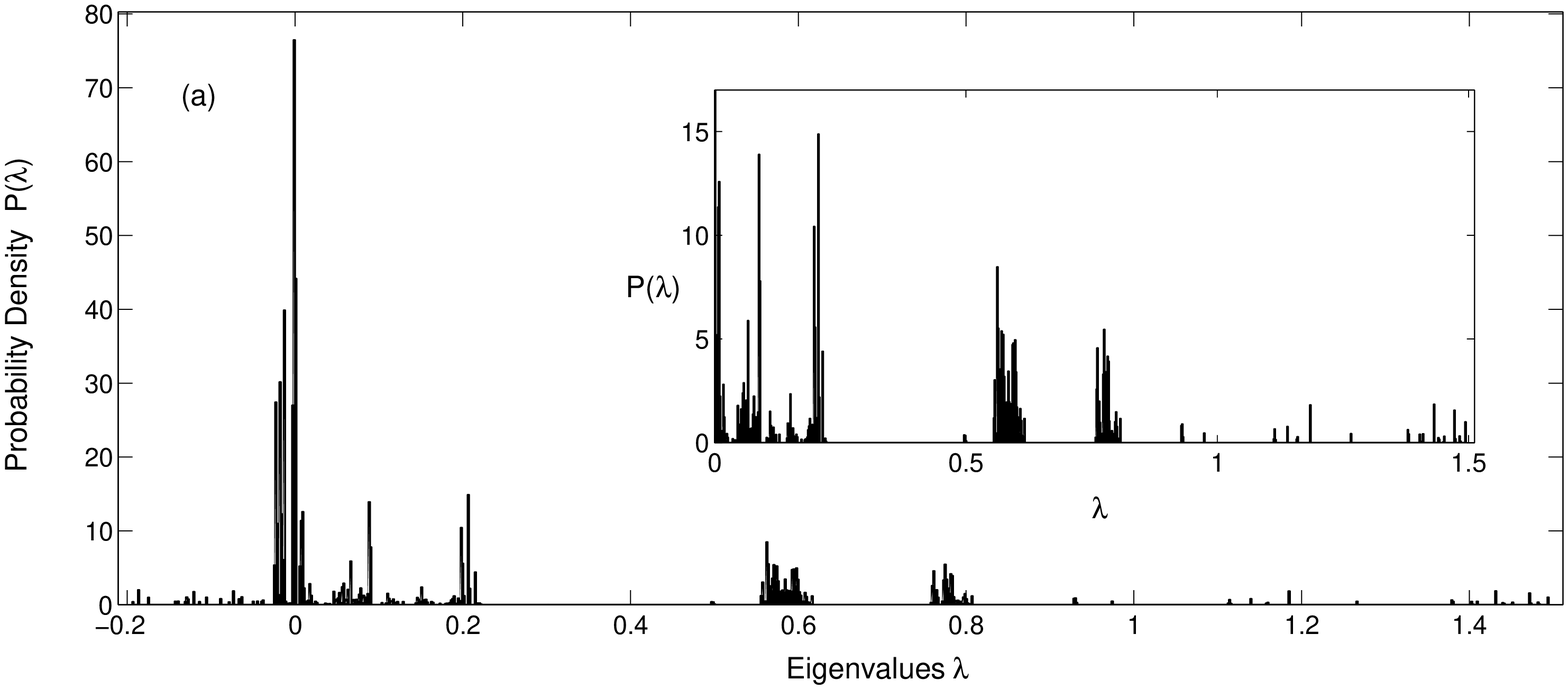}
\includegraphics[width=70mm,height=70mm]{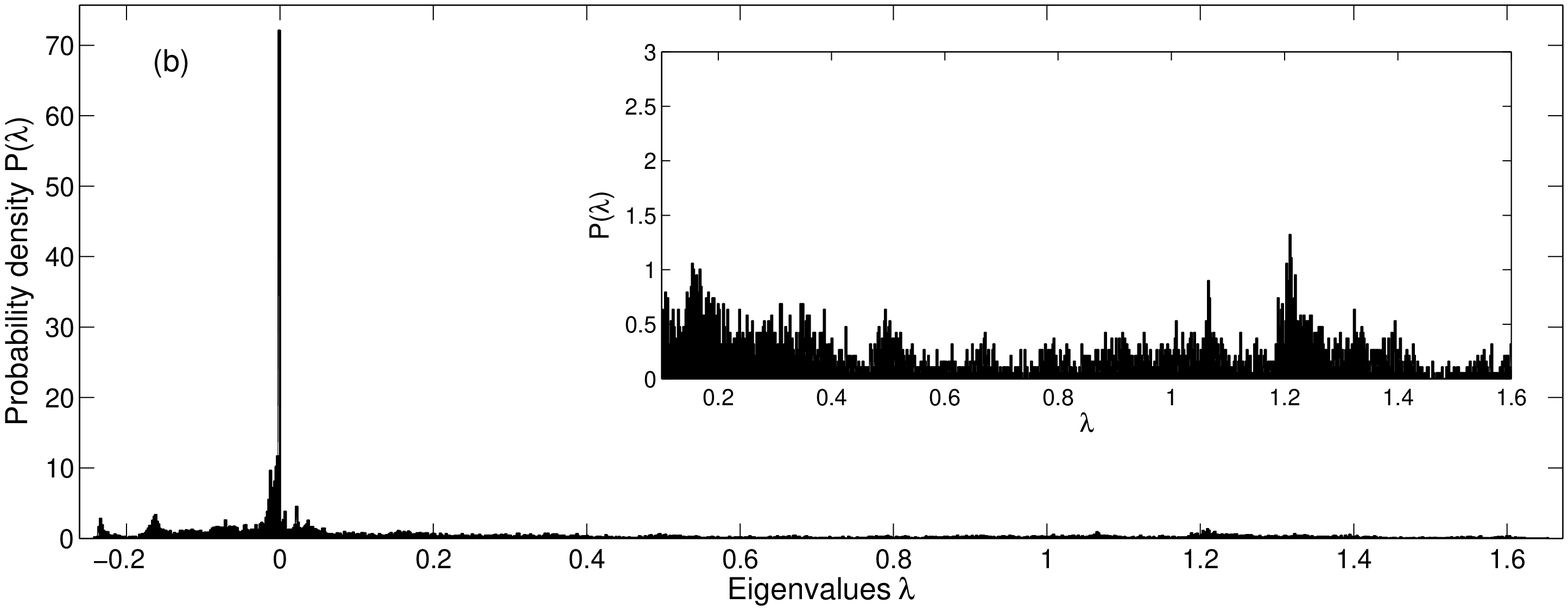}
\caption{\label{fig9} The eigenvalue distribution for (a) $\epsilon = 0.2$, (b) 
$\epsilon = 0.3$ is plotted for 50 different realizations for lattice size 
$L = 1000$ with bin-width = 0.001. Gaps are seen in localized chaos eigenvalue 
distribution whereas the eigenvalue distribution for Spatiotemporal chaos region
shows no such gaps. The inset in figure (a),(b) show the magnified region of respective 
plots}.
\end{figure*}

Fig. \ref{fig9}(a) shows that the eigenvalue distribution for $\epsilon = 0.2$, where 
spatial intermittency or localized chaos is seen. This distribution has gap or zero probability 
regions in the eigenvalue spectrum as reported earlier by Jabeen {\it {et al.}} \cite{gupte74}.
 On the other hand, Fig. \ref{fig9}(b) shows that there is no such gap for coupling values $\epsilon = 0.3$
 where fully grown spatiotemporal chaos is seen. 
Consider matrix $J_t'$ given by
\begin{displaymath}
\mathbf{J'_t} =
\left(\begin{array}{cc}
A'_t  & B'_N \\
B'_N & 0_N
\end{array}\right)
\end{displaymath}
where $B'(i,i) = \sqrt{\delta f'(y_i(t))}$
and $B'(i,j) = 0$,
 if $i\neq j$. The elements of the matrix $A'$ are $A'(i,i) = A(i,i)$ and 
$A'(i,j)=\sqrt{A(i,j)A(j,i)}$ which can be obtained from the transformation $A'(t)=C'AC'^{-1}$ where 
matrix $C'(i,i)=\sqrt{\epsilon_n f'(x_i(t))}$.

One can verify that $B'^2=B$. Thus, it is easy to see that matrix $J'(t)$ and matrix $J(t)$ have same characteristic 
polynomial and hence they have same eigenvalues.  
$\displaystyle  \left( \begin{array}{cc} C' & 0 \\ 
0 & C'B' \end{array}\right)$ is the similarity transform which relates $J_t$ and $J'_t$.
Now the matrix $J'(t)$ is symmetric and has real eigenvalues.
Thus, the matrix $J(t)$ also should have  real eigenvalues.  The assumption has been that
 $f'(x_i) \geq 0,  f'(y_i) \geq 0$ which is justified for $k \leq 1$. The feedback strength
$\delta$ is already assumed to be positive in our work. Thus $A(t)$ and $A'(t)$ as well 
as $J_t$ and $J'_t$ are related by similarity transformation and since $A'(t)$ and $J_t'$ are symmetric,
$A(t)$ and $J_t$ should have real eigenvalues. Thus we can look for gap
in eigenvalue spectrum for these matrices and we observe the gap in our case
for $\tau=1$. Gupte and Jabeen already noticed the gap for case without feedback.
We would also like to note that these are stretching transformations and
eigenvectors which are localized in one frame of reference remain
localized in other frame of reference as well. This observation will
be useful in later section. Thus it is possible to look for gap in the 
eigenvalue spectrum even in this case. Unfortunately the argument breaks down for $\tau>1$ \cite{footnote3}.

Now let us try to understand the reason for gap in eigenvalue spectrum in $A(t)$ in Jabeen and Gupte's case.
We choose a diagonal part of matrix as an unperturbed matrix and all off-diagonal entries as a perturbation. Our choice 
for $l^{th}$ eigenvector  for the diagonal part is a vector whose $l^{th}$ component is unity 
and all other components are zero. We define $A'_t = A^d_t +A^p_t$, where diagonal entries of $A^d_t$
 are given by $A^d_t(i,i)=A_t(i,i)$ and rest of its entries are zero. Let us consider an idealized case 
in which $k^{th}$ and $N^{th}$ sites are turbulent and all other sites are laminar. Similar
analysis can be carried out for more that two turbulent sites as long as they are not adjacent sites.
\begin{displaymath}
\mathbf{A'_t} =
\left(\begin{array}{cccccccccc}
a & b & 0    & 0   & 0 & 0 & 0 & \ldots & 0 & b_N \\
b & a & b    & 0   & 0 & 0 & 0 & \ldots & 0 & 0 \\
\vdots & \vdots & \ddots    & \vdots   & \vdots  & \vdots & \vdots & \ldots & 0 & 0 \\
0 & 0 & \ldots & a  & b_k & 0 & 0  & \ldots  & 0 & 0\\
0 & 0 & \ldots & b_k  & a_k & b_k & 0 & \ldots & 0 & 0\\
0 & 0 & \ldots & 0    & b_k & a  & b & \ldots & 0 & 0\\
\vdots & \vdots & \vdots & \vdots & \vdots & \vdots & \vdots & \ddots& \vdots& \vdots\\
0 & 0 & 0    & 0   &0  & 0 & 0& \ldots  & a & b_N\\
b_N & 0 & 0    & 0   &0 & 0 & 0& \ldots  & b_N & a_N\\
\end{array}\right)
\end{displaymath}

where $a= (1-\epsilon) f'(x^*)$, $b= \epsilon_n f'(x^*)$ and $a_i=(1-\epsilon) f'(x_i)$, $b_i= \epsilon_n \sqrt{f'(x_i)f'(x^*)}$, 
where $i = k$ and $N$ sites respectively. The matrix $A^d_t$ has eigenvalue 
$a$ with $(N-2)$ fold degeneracy and two nondegenerate eigenvalues given by $a_k$ and $a_N$.
The perturbation is $A^p_t=A'_t-A^d_t$. If all sites are laminar except $k^{th}$ and $N^{th}$ site 
(since we have periodic boundary conditions, by appropriate relabeling of sites, one can always bring 
the site in turbulent state to $N^{th}$ site), the matrix $A^d_t=diag(a,a, a_k\ldots ,a ,a_N)$ , the perturbation will be,
\begin{displaymath}
\mathbf{A^p_t} =
\left(\begin{array}{cccccccccc}
0 & b & 0    & 0   & 0 & 0 & 0 & \ldots & 0 & b_N \\
b & 0 & b    & 0   & 0 & 0 & 0 & \ldots & 0 & 0 \\
\vdots & \vdots & \ddots    & \vdots   & \vdots  & \vdots & \vdots & \ldots & 0 & 0 \\
0 & 0 & \ldots & 0  & b_k & 0 & 0  & \ldots  & 0 & 0\\
0 & 0 & \ldots & b_k  & 0 & b_k & 0 & \ldots & 0 & 0\\
0 & 0 & \ldots & 0    & b_k & 0  & b & \ldots & 0 & 0\\
\vdots & \vdots & \vdots & \vdots & \vdots & \vdots & \vdots & \ddots& \vdots& \vdots\\
0 & 0 & 0    & 0   &0  & 0 & 0& \ldots  & 0 & b_N\\
b_N & 0 & 0    & 0   &0 & 0 & 0& \ldots  & b_N & 0\\
\end{array}\right)
\end{displaymath}

We perform a simple degenerate state perturbation theory to our perturbation matrix
 $A^p_t$, we can consider a matrix $P$ of $(N-2)$ eigenvectors of degenerate eigenvalues as,
\begin{displaymath}
\mathbf{P} = 
\left( \begin{array}{cccccccc} 
1      &    0   &   0    &   0    &   0    &   0    &   0    & 0\\
0      &    1   &   0    &   0    &   0    &   0    &   0    & 0\\
\vdots & \vdots & \ddots & \vdots & \vdots & \vdots & \vdots & \vdots \\
0      & 0	&   0	 &   1    &   0    &   0    &   0    & 0 \\
0      & 0	&   0 	 &   0    &   0    &   0    &   0    & 0 \\
0      & 0	&   0    &   0    &   1    &   0    &   0    & 0 \\
0      & 0	&   0    &   0    &   0    &   1    &   0    & 0 \\
\vdots & \vdots & \vdots & \vdots & \vdots & \vdots & \ddots & \vdots\\
0      & 0	&   0    &   0    &   0    &   0    &   0    & 1\\
0      & 0	&   0    &   0    &   0    &   0    &   0    & 0\\
\end{array}\right)
\end{displaymath} 

it can be seen that
\begin{displaymath}
\mathbf{P^T A^p_t P} =
\left( \begin{array}{ccccccc|ccccccc} 
0 & b & 0 & 0  & \ldots & 0 & 0 & 0 & 0 & 0 & 0 &\ldots & 0\\
b & 0 & b & 0  & \ldots & 0 & 0 & 0 & 0 & 0 & 0 &\ldots & 0\\
0 & b & 0 & b  & \ldots & 0 & 0 & 0 & 0 & 0 & 0 &\ldots & 0\\
0 & 0 & b & 0  & \ddots & 0 & 0 & 0 & 0 & 0 & 0 &\ldots & 0\\
\vdots & \vdots & \vdots & \ddots & \ddots & \vdots & \vdots &  \vdots & \vdots & \vdots & \vdots & \ldots &\vdots\\
0 & 0 & 0 & 0  &\ldots & 0 & b & 0& 0 & 0 & 0 & \ldots & 0\\
0 & 0 & 0 & 0  &\ldots & b & 0 & 0& 0 & 0 & 0 & \ldots & 0\\
\hline
0 & 0 & 0 & 0  & \ldots & 0 & 0 & 0& b & 0 & 0 & \ldots & 0\\
0 & 0 & 0 & 0  & \ldots & 0 & 0 & b& 0 & b & 0 & \ldots & 0\\
0 & 0 & 0 & 0  & \ldots & 0 & 0 & 0& b & 0 & b & \ldots & 0\\
0 & 0 & 0 & 0  & \ldots & 0 & 0 & 0& 0 & b & 0 & \ddots & 0\\
\vdots & \vdots & \vdots & \vdots & \vdots & \vdots & \vdots & \vdots & \vdots & \vdots & \ddots & \ddots & b\\
0 & 0 & 0 & 0  & \ldots & 0 & 0 & 0& 0 & 0 & 0 & b & 0\\
\end{array}\right)
\end{displaymath}
\noindent

This gives us for each $(N-2)$ fold degenerate eigenvector, a $(N-2)\times(N-2)$ matrix of perturbation $A^p_t$.
We can see that the above matrix contains two blocks. The first block is of $(k-1) \times (k-1)$ size and the second block has size
$(N-k-1) \times (N-k-1)$.
The eigenvalues of this matrix will be first order corrections. The perturbed 
eigenvalues after first order perturbation for degenerate states in this case will 
be,
\begin{equation}
\label{eqn7}
\lambda_l = a+2b\cos(\theta_l)
\end{equation}

 where \cite{toeplitz},

\begin{equation}
 \theta_l = \left\{
\begin{array} {lll}
  \frac{l \pi}{(k-1)+1}       &  \text{for }  l = 1,..,k-1 \\
  \frac{l \pi}{(N-k-1)+1} &  \text{for }  l = 1,..,(N-k-1)
\end{array}\right.
\end{equation} 
\noindent
We have two non-degenerate eigenvectors, an unit vector in $N^{th}$ direction 
$v_N= [0,0,\ldots ,1]^T$ and an unit vector in $k^{th}$ direction 
$v_k=[0,0,\ldots 1, \ldots, 0]^T$.  Now $v_N^T A_t^p v_N=v_k^T A_t^p v_k =0$. 
Thus the eigenvalues $a_k$ and $a_N$ are unperturbed after first order perturbation. 
Now if these values are far from $a \pm 2b$, we will observe a gap in the spectrum.

Now let us generalize this argument to finite fraction $f$ of turbulent sites. 
The number of turbulent sites is $m=fN$ and they are interspersed in the lattice. 
We also assume that no two turbulent sites are adjacent.
In that case, we have $a_i=(1-\epsilon) f'(x_i)$, $b_i=\epsilon_n(f'(x_i))$,
 where $i = l_1, l_2, l_3,...,l_m$ sites respectively, where $l_1,l_2,..,l_m$ are indices of $m$ turbulent 
sites and $l_m$ site is $N^{th}$ site which can be obtained due to 
periodic boundary conditions. Again the matrix $A^d_t$ will have eigenvalue
 $a$ with $(N-m)$ fold degeneracy and $m$ nondegenerate eigenvalues given by $a_1, a_2,...,a_m$. The matrix 
$P$ becomes matrix of $(N-m)$ eigenvectors of degenerate eigenvalues. The matrix of 
perturbation $A^p_t$ changes accordingly to $(N-m)\times(N-m)$. We can see that the matrix obtained 
from operation $P^T A^p_t P$ contains $m$ blocks. The square blocks will be of corresponding size, i. e.
for $l_1$, the size will be  $(l_1-1)$, for $l_2$ it is $(l_2-l_1-1)$ and in the same way for 
$m^{th}$ block corresponding to turbulent site $l_m$, the size will be $(l_m-l_{(m-1)}-1)$. 
The generalized case for $m$ turbulent sites goes as,
\begin{equation}
\label{eqn9}
 \theta_l = \left\{
\begin{array} {lll}
  \frac{l \pi}{l_1}       &  \text{for }  l = 1,..,l_1-1 \\
  \frac{l \pi}{l_2-l_1}       &  \text{for }  l = 1,..,l_2-l_1-1 \\
  \frac{l \pi}{l_3-l_2}       &  \text{for }  l = 1,..,l_3-l_2-1 \\
  \frac{l \pi}{l_4-l_3}       &  \text{for }  l = 1,..,l_4-l_3-1 \\
  \frac{l \pi}{l_m-l_{(m-1)}}       &  \text{for }  l = 1,..,l_m-l_{m-1}-1 \\
\end{array}\right.
\end{equation} 

\begin{figure}
\includegraphics[width=60mm,height=60mm]{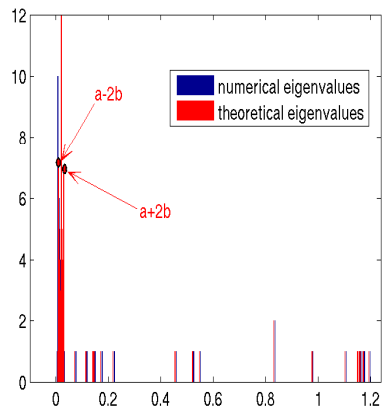}
\caption{\label{fig10} (Color online) Overlapping histograms of the calculated eigenvalues. In red (gray), 
histogram of eigenvalues obtained by theory of pertubative analysis. In blue(black), histogram of 
numerically obtained eigenvalues of matrix $A(t)$ is given. The values for turbulent sites given 
by $\epsilon_n f'(x_j))$ can also be seen away from the band $[a-2b,a+2b]$. The agreement between 
theory and numerics can be clearly seen for degenerate as well as non-degenerate eigenvalues.}
\end{figure}

\begin{figure}
\includegraphics[width=60mm,height=60mm]{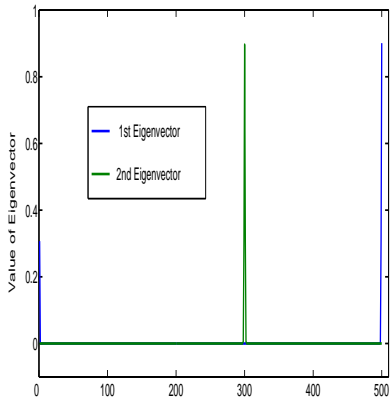}
\caption{\label{fig11} The localized eigenvectors showing peaks at $300$ and $500$ for lattice size $500$. }.
\end{figure}

In any case the eigenvalues are between $[a-2b,a+2b]$. Now let us look at nondegenerate 
states. Arguing similar to the case of two site, after first order perturbation, 
the eigenvalues are $a_{l_1}, a_{l_2},\ldots a_{l_m}$. If these eigenvalues are not 
within the range $a \pm 2b$, we will observe a gap in the spectrum. 
Thus, we predict that eigenspectrum of
$J_t$ will be given by a band of eigenvalues between $a+2b$ and $a-2b$ superposed with
values $\epsilon_n(f'(x_j))$, where $j$ is a turbulent site. Localized eigenstates for 
$A'(t)$ will also be localized for $A(t)$ since they are related by stretching transformation.
If we make the histogram of eigenvalues as shown in figure \ref{fig10}, we find that this 
is indeed true. We can see histograms of eigenvalues computed numerically and theoretically from 
the eqns. \ref{eqn7} and \ref{eqn9} obtained by pertubative analysis for $20$ turbulent 
sites in the lattice of $100$ elements. The values for turbulent sites given by 
$\epsilon_n f'(x_j))$ can also be seen away from the band $[a-2b,a+2b]$. For general case, there is a 
band of eigenvalues followed by $m$ non-degenerate eigenstates which are far from the band. 
We checked that these eigenstates are localized and the corresponding 
eigenvectors are demonstrated in Fig. \ref{fig11} for case of two turbulent sites.

On the other hand, for spatiotemporal chaos, we have a symmetric random matrix $A'_t$ which has nonzero entries on diagonal,
upper diagonal, lower diagonal and elements $A'_t(1,N)$ and $A'_t(N,1)$. In this case, ratio of width of band
to system size tends to zero as $N \rightarrow \infty$. The eigenvalue spectrum of such a matrix tends
to standard Gaussian distribution \cite{liuwang}. It is worth noting here that, persistence acts 
as a good order parameter for this transition for all values of $\tau$ including $\tau=0$, 
whereas it is not clear that dynamic characterizer of gaps in eigenvalues spectrum might work for higher value of $\tau$. 
 \begin{table}
\caption{ Table showing Eigenvectors showing peaks at corresponding two turbulent sites in otherwise steady state, here $300$ and $500$. In second column larger eigenvalues are given which match with the values shown in last column.}
 \label{table1} 
\centering
 \begin{tabular}{|c|c|c|c|}
  \hline \hline
  $i$ & Eigenvector & Eigenvalue & $(1-\epsilon) f'(x)$\\
  \hline 
  500 & 1 & 1.0861 &  1.0817\\
  300 & 2 & 0.4177 &  0.4132 \\
\hline \hline
 \end{tabular}
 \end{table}


Similar calculations can be carried out for our Jacobian $J'_t$. The localized eigenvectors
 explain why localized patterns are stable. The eigenvectors having largest eigenvalue are 
localized. Thus any perturbation from laminar state is likely to be growing
in the direction of localized eigenvectors which does not disturb values at nearby
 sites. The above analysis of eigenvalue distribution for localized chaos is restricted to
 demonstrate the analogy to spatial intermittency. The eigenvalue spectrum clearly 
 differentiates infective to non-infective phase. The above analysis explains why the eigenvalue spectrum is
different for spatiotemporal chaos and spatial intermittency. \cite{footnote4}

\section{Discussion}
 
In this paper, we have simulated  coupled circle maps with feedback. This allows us to 
study coupled systems where local dynamics is high dimensional. 
We give an overall phase diagram and discuss  various dynamical phases.
We also give an all-site bifurcation diagram and remark that there
are phases which can not be found from bifurcation diagram and we need 
to study a detailed simulated space-time plot. In particular, we 
make a detailed study of a dynamical phase transition between the phase
of localized chaos and spatiotemporally chaotic state. We find various
characteristics such as distribution of domains of turbulent sites. 
Phase transition from different dynamical phases is demonstrated. We make
 an attempt to analyze this transition in a manner in which equilibrium 
transitions are analyzed. In order that we are able to do so, it is necessary
 to define an order parameter. We find that local persistence acts as 
an excellent order parameter to characterize the phase transition. In 
addition to this, conventional scaling laws pertaining to finite size 
simulations and offcritical scaling add credibility to use of persistence as 
quantifier of the transition. Interestingly, same definition of persistence used 
for without feedback case \cite{menon} works for our case with feedback.  We also calculate
one step Jacobian of the system and find gaps in eigenvalue distribution
of system which is dynamical signature of non-infectious nature of dynamics
in `localized chaos' regime. The important effect of feedback is that the parameter window
for which spatial intermittency is seen, gets enhanced. It should be noted that 
persistence acts as a good order parameter for this transition for all values 
of $\tau$ including $\tau=0$, whereas it is not clear that dynamic characterizer 
of gaps in eigenvalues spectrum might work for higher value of $\tau$. 
We get this transition for several values of $\tau$. We have shown
that persistence acts as an excellent order parameter for this
transition for all values of $\tau$. We also try to study gaps in the
eigenvalue spectrum which work well for $\tau=0$ and $\tau=1$.
We go a step ahead in explaining the origin
of such gaps in eigenvalues spectra of one step Jacobian by perturbation
theory analysis. Analytical work also gives an idea of when above analysis may not work.

There is a possibility to experimentally investigate dynamical phases
in spatiotemporal systems. Certain electrical oscillators are well simulated with circle 
maps \cite{afraimovich,kurths} and it is relatively easy to implement feedback
 mechanism in electric devices. If one can define a persistence in an 
analogous manner in differential equations, it is possible to find 
persistence in spatiotemporal systems experimentally.

In context of localized chaos, we would also like to point to
an interesting work by Sethia et. al. \cite{Sethia} where they observe 
`chimera' states, first observed by Kuramoto \cite{kuramoto}. These states have 
spatially disconnected regions of synchronization separated by regions of decoherence. 
Thus there is a strong analogy between `chimera' states and `localized chaos'. 
Several works study `chimera' under various modifications for identical 
oscillators.\cite{abrams,tass,sheeba,wiley} and non-identical oscillators \cite{laing}.
Coupled evolution of systems with delay induces effectively nonlocal contributions in the 
evolution of individual system. Feedback propagates in space and time and affects 
the dynamics at some distant site. It could be of interest to find some general 
definition of persistence applicable to these systems and investigate transition 
to `chimera' state as a dynamic phase transition. Further work in this direction is pursued.

ARS would like to thank D.A.E, India and C.S.I.R, India for financial support and
 Dr. S. Barve for useful discussions. PMG would like to thank DST for financial support.



\begin{thebibliography}{24}
\expandafter\ifx\csname natexlab\endcsname\relax\def\natexlab#1{#1}\fi
\expandafter\ifx\csname bibnamefont\endcsname\relax
  \def\bibnamefont#1{#1}\fi
\expandafter\ifx\csname bibfnamefont\endcsname\relax
  \def\bibfnamefont#1{#1}\fi
\expandafter\ifx\csname citenamefont\endcsname\relax
  \def\citenamefont#1{#1}\fi
\expandafter\ifx\csname url\endcsname\relax
  \def\url#1{\texttt{#1}}\fi
\expandafter\ifx\csname urlprefix\endcsname\relax\def\urlprefix{URL }\fi
\providecommand{\bibinfo}[2]{#2}
\providecommand{\eprint}[2][]{\url{#2}}



\bibitem[{\citenamefont{biology}(1991)}]{biology}
\bibinfo{author}{\bibfnamefont{A. J.}~\bibnamefont{Koch}} and
\bibinfo{author}{\bibfnamefont{H.}~\bibnamefont{Meinhardt}},
  \bibinfo{journal}{Rev. Mod. Phys.} \textbf{\bibinfo{volume}{66}},
  \bibinfo{pages}{1481} (\bibinfo{year}{1991}).


\bibitem[{\citenamefont{JJA}(1988)}]{JJA}
\bibinfo{author}{\bibfnamefont{K.}~\bibnamefont{Wiesenfeld,}} 
\bibinfo{author}{\bibfnamefont{P.}~\bibnamefont{Hadley}},
  \bibinfo{journal}{Phys. Rev. Lett} \textbf{\bibinfo{volume}{62}},
  \bibinfo{pages}{1335} (\bibinfo{year}{1988}).
\bibitem{bak} Bak et al.{ Solid State Commun.} {\bf 51,}(1984) 231.
\bibitem[{\citenamefont{Joseph}(1996)}]{Joseph} 
\bibinfo{author}{\bibfnamefont{K.}~\bibnamefont{Wiesenfeld.}} 
  \bibinfo{journal}{Physica (Amsterdam)} \textbf{\bibinfo{volume}{222(B)}},
  \bibinfo{pages}{315} (\bibinfo{year}{1996}).
\bibitem[{\citenamefont{rroy}(2007)}]{rroy}
\bibinfo{author}{\bibfnamefont{L.}~\bibnamefont{Illing,}} 
\bibinfo{author}{\bibfnamefont{D. J.}~\bibnamefont{Gauthier,}}
\bibinfo{author}{\bibfnamefont{R.}~\bibnamefont{Roy,}}  
  \bibinfo{journal}{Adv. in Atomic, Molecular and Opt. Phys.}
 \textbf{\bibinfo{volume}{54}},
  \bibinfo{pages}{615} (\bibinfo{year}{2007}).
\bibitem[{\citenamefont{cross}(1993)}]{cross}
\bibinfo{author}{\bibfnamefont{M. C.}~\bibnamefont{Cross,}} 
\bibinfo{author}{\bibfnamefont{P. C.}~\bibnamefont{Hohenberg,}} 
  \bibinfo{journal}{Rev. Mod. Phys.} \textbf{\bibinfo{volume}{65}},
  \bibinfo{pages}{851} (\bibinfo{year}{1993}) and references therein.
\bibitem[()]{footnote1}
\bibinfo{note}{One could also add cellular automata to 
these list of models. However, we feel that due to discreteness in the value of 
variables, it displays very different patterns. Usually there is no tunable
 parameter and bifurcations can not be studied. However, 
certain patterns in a given cellular automata could look very similar to certain
 coarse grained patterns in continuous systems}
\bibitem[{\citenamefont{chate}(1988)}]{chate} 
\bibinfo{author}{\bibfnamefont{H.}~\bibnamefont{Chate,}} 
\bibinfo{author}{\bibfnamefont{P.}~\bibnamefont{Manneville.}}
  \bibinfo{journal}{Physica D} \textbf{\bibinfo{volume}{32}},
  \bibinfo{pages}{409} (\bibinfo{year}{1988}).
\bibitem[{\citenamefont{bigazzi}(1988)}]{bigazzi}
\bibinfo{author}{\bibfnamefont{S.}~\bibnamefont{Ciliberto,}} 
\bibinfo{author}{\bibfnamefont{P.}~\bibnamefont{Bigazzi}},
  \bibinfo{journal}{Phys. Rev. Lett} \textbf{\bibinfo{volume}{60}},
  \bibinfo{pages}{286} (\bibinfo{year}{1988}).
\bibitem[{\citenamefont{barkley}(1990)}]{barkley}
\bibinfo{author}{\bibfnamefont{D.}~\bibnamefont{Barkley}},
  \bibinfo{title}{ {\em{Nonlinear Structures in Dynamical Systems}}, 
edited by Lui Lam and H. C. Moris }
  (\bibinfo{publisher}{Springer-Verlag}, \bibinfo{address}{New York},
  \bibinfo{year}{1990}).
\bibitem[{\citenamefont{reynolds}(1990)}]{reynolds}
\bibinfo{author}{\bibfnamefont{D. A.} \bibnamefont{Kessler,}}
   \bibinfo{author}{\bibfnamefont{H.}~\bibnamefont{Levine,}}
\bibinfo{author}{\bibfnamefont{W. N.} \bibnamefont{Reynolds,}}
\bibinfo{journal}{Phys. Rev. A} \textbf{\bibinfo{volume}{42}},
  \bibinfo{pages}{6125} (\bibinfo{year}{1990}).
\bibitem[{\citenamefont{jampa}(2007)}]{jampa}
\bibinfo{author}{\bibfnamefont{M. P. K} \bibnamefont{Jampa,}}
   \bibinfo{author}{\bibfnamefont{A. R.}~\bibnamefont{Sonawane,}}
\bibinfo{author}{\bibfnamefont{P. M.} \bibnamefont{Gade,}}
\bibinfo{author}{\bibfnamefont{S.} \bibnamefont{Sinha}},
\bibinfo{journal}{Phys. Rev. E} \textbf{\bibinfo{volume}{75}},
  \bibinfo{pages}{026215} (\bibinfo{year}{2007}).
\bibitem[{\citenamefont{politi_chaos92}(1992)}]{politi_chaos92}
\bibinfo{author}{\bibfnamefont{A.} \bibnamefont{Politi,}}
   \bibinfo{author}{\bibfnamefont{A.}~\bibnamefont{Torcini,}}
\bibinfo{journal}{Chaos} \textbf{\bibinfo{volume}{2(3)}},
  \bibinfo{pages}{293} (\bibinfo{year}{1992}).
\bibitem[{\citenamefont{bonetto_CUP}(2005)}]{bonetto_CUP}
\bibinfo{author}{\bibfnamefont{F.} \bibnamefont{Bonetto,}}
   \bibinfo{author}{\bibfnamefont{A.}~\bibnamefont{Kupiainen,}}
\bibinfo{author}{\bibfnamefont{J. L.} \bibnamefont{Lebowitz,}}
\bibinfo{journal}{Ergod. Th. \& Dynam. Sys.} \textbf{\bibinfo{volume}{25}},
  \bibinfo{pages}{59} (\bibinfo{year}{2005})

\bibitem[{\citenamefont{sakaguchi_pre99}(1999)}]{sakaguchi_pre99} 
\bibinfo{author}{\bibfnamefont{H.}~\bibnamefont{Sakaguchi.}} 
  \bibinfo{journal}{Phys. Rev. E)} \textbf{\bibinfo{volume}{60}},
  \bibinfo{pages}{7584} (\bibinfo{year}{1999}).
\bibitem[{\citenamefont{kaneko}(1993)}]{kaneko}
\bibinfo{author}{\bibfnamefont{K.}~\bibnamefont{Kaneko}},
  \bibinfo{title}{{\em{Theory and Applications of Coupled Map Lattices}}}
  (\bibinfo{publisher}{Wiley}, \bibinfo{address}{New York},
  \bibinfo{year}{1993}).
\bibitem[{\citenamefont{stanley}(1995)}]{stanley}
\bibinfo{author}{\bibfnamefont{A. L.}~\bibnamefont{Barabasi, }} 
\bibinfo{author}{\bibfnamefont{H.}~\bibnamefont{Stanley}}, 
  \bibinfo{title}{ {\em{Fractal Concepts in Surface Growth}}},
  (\bibinfo{publisher}{Cambridge University Press},
\bibinfo{address}{U.K},\bibinfo{year}{1995})
\bibitem[{\citenamefont{wolf}(1996)}]{wolf}
\bibinfo{author}{\bibfnamefont{D. E.}~\bibnamefont{Wolf,}} 
\bibinfo{author}{\bibfnamefont{M.}~\bibnamefont{Schreckenberg and}} 
\bibinfo{author}{\bibfnamefont{A.}~\bibnamefont{Bachem,}} 
  \bibinfo{title}{ {\em{Traffic and Granular Flow}}},
  \bibinfo{publisher}{World Scientific,}
\bibinfo{address}{Singapore},\bibinfo{year}{1996}
\bibitem[{\citenamefont{haye}(2000)}]{haye}
\bibinfo{author}{\bibfnamefont{H.}~\bibnamefont{Hinrichson,}} 
  \bibinfo{journal}{Advances in Physics} \textbf{\bibinfo{volume}{49}},
  \bibinfo{pages}{815-958} (\bibinfo{year}{2000}).
\bibitem[{\citenamefont{gupte72}(2005)}]{gupte72}
\bibinfo{author}{\bibfnamefont{Z.} \bibnamefont{Jabeen}} \bibnamefont{and}
\bibinfo{author}{\bibfnamefont{N.} \bibnamefont{Gupte}},
  \bibinfo{journal}{Phys. Rev. E} \textbf{\bibinfo{volume}{72}},
  \bibinfo{pages}{016202} (\bibinfo{year}{2005}).
\bibitem[{\citenamefont{gupte74}(2006)}]{gupte74}
\bibinfo{author}{\bibfnamefont{Z.}~\bibnamefont{Jabeen}} \bibnamefont{and}
  \bibinfo{author}{\bibfnamefont{N.}~\bibnamefont{Gupte}},
  \bibinfo{journal}{Phys. Rev. E} \textbf{\bibinfo{volume}{74}},
  \bibinfo{pages}{016210} (\bibinfo{year}{2006}).
\bibitem[{\citenamefont{rahulpandit}(2000)}]{rahulpandit}
\bibinfo{author}{\bibfnamefont{A.}~\bibnamefont{Pande}} \bibnamefont{and}
  \bibinfo{author}{\bibfnamefont{R.}~\bibnamefont{Pandit}},
  \bibinfo{journal}{Phys. Rev. E} \textbf{\bibinfo{volume}{61}},
  \bibinfo{pages}{6448} (\bibinfo{year}{2000}).

\bibitem[{\citenamefont{gade}(2007)}]{gade}
\bibinfo{author}{\bibfnamefont{P.M.} \bibnamefont{Gade,}}
\bibinfo{author}{\bibfnamefont{D.V.} \bibnamefont{Senthilkumar,}}
   \bibinfo{author}{\bibfnamefont{S.}~\bibnamefont{Barve,}}
\bibinfo{author}{\bibfnamefont{S.} \bibnamefont{Sinha}},
  \bibinfo{journal}{Phys. Rev. E} \textbf{\bibinfo{volume}{75}},
  \bibinfo{pages}{066208} (\bibinfo{year}{2007}).
\bibitem[{\citenamefont{ashwini}(2010)}]{ashwini}
\bibinfo{author}{\bibfnamefont{A. V.} \bibnamefont{Mahajan,}}
   \bibinfo{author}{\bibfnamefont{P. M.}~\bibnamefont{Gade,}}
  \bibinfo{journal}{Phys. Rev. E} \textbf{\bibinfo{volume}{81}},
  \bibinfo{pages}{056211} (\bibinfo{year}{2010}).

\bibitem[{\citenamefont{ginelli}(2003)}]{ginelli}
\bibinfo{author}{\bibfnamefont{F.} \bibnamefont{Ginelli,}}
\bibinfo{author}{\bibfnamefont{R.} \bibnamefont{Livi,}}
   \bibinfo{author}{\bibfnamefont{A.}~\bibnamefont{Politi,}}
\bibinfo{author}{\bibfnamefont{A.} \bibnamefont{Torcini}},
  \bibinfo{journal}{Phys. Rev. E} \textbf{\bibinfo{volume}{67}},
  \bibinfo{pages}{046217} (\bibinfo{year}{2003}).

\bibitem[{\citenamefont{Pyragas}(1992)}]{pyragas}
  \bibinfo{author}{\bibfnamefont{K.}~\bibnamefont{Pyragas}},
  \bibinfo{journal}{Phys. Lett. A} \textbf{\bibinfo{volume}{170(6)}},
  \bibinfo{pages}{421-428} (\bibinfo{year}{1992}).
\bibitem[{\citenamefont{handbook}(1993)}]{handbook}
\bibinfo{author}{\bibfnamefont{H.}~\bibnamefont{Schuster}},
  \emph{\bibinfo{title}{Handbook of Chaos Control}}
  (\bibinfo{publisher}{Wiley-VCH}, \bibinfo{address}{Weinheim},
  \bibinfo{year}{1999}).

\bibitem[{\citenamefont{frozen}(1993)}]{frozen}
\bibinfo{author}{\bibfnamefont{F. H.} \bibnamefont{Willeboordse}},
  \bibinfo{journal}{Phys. Lett. A} \textbf{\bibinfo{volume}{183}},
  \bibinfo{pages}{187-192} (\bibinfo{year}{1993}).
\bibitem[{\citenamefont{menon}(2003)}]{menon}
\bibinfo{author}{\bibfnamefont{G.I.} \bibnamefont{Menon,}}
   \bibinfo{author}{\bibfnamefont{S.}~\bibnamefont{Sinha,}}
\bibinfo{author}{\bibfnamefont{P.C.} \bibnamefont{Ray}},
  \bibinfo{journal}{Europhys. Lett.} \textbf{\bibinfo{volume}{61(1)}},
  \bibinfo{pages}{27-33} (\bibinfo{year}{2003}).

\bibitem[{\citenamefont{abhijeet_dae}(2006)}]{abhijeet_dae}
\bibinfo{author}{\bibfnamefont{A. R.} \bibnamefont{Sonawane,}}
   \bibinfo{author}{\bibfnamefont{P. M.}~\bibnamefont{Gade,}}
  \bibinfo{journal}{Proc. of DAE Solid State Physics Symposium} \textbf{\bibinfo{volume}{51}},
  \bibinfo{pages}{129} (\bibinfo{year}{2006}).


\bibitem[{\citenamefont{satya}(1999)}]{satya}
\bibinfo{author}{\bibfnamefont{S.~N.} \bibnamefont{Majumdar}},
  \bibinfo{journal}{Curr. Sci.} \textbf{\bibinfo{volume}{77}},
  \bibinfo{pages}{370} (\bibinfo{year}{1999}).
\bibitem[{\citenamefont{koduvely}(1998)}]{koduvely}
\bibinfo{author}{\bibfnamefont{H.} \bibnamefont{Hinrichsen}},
\bibinfo{author}{\bibfnamefont{H. M.} \bibnamefont{Koduvely,}}
  \bibinfo{journal}{Eur. Phys. J. B} \textbf{\bibinfo{volume}{5}},
  \bibinfo{pages}{257-264} (\bibinfo{year}{1998}).
\bibitem[()]{footnote2}
\bibinfo{note}{
We note that if the number of laminar sites are few and far between which will 
happen at parameter values extremely close to  the transition point, persistence 
may still go to zero asymptotically since even a small fluctuation will make a 
non-persistent site persistent. Still, it is fair to say that the critical value
$\epsilon_c$ at which a power law decay is observed, is very close to, if not
the same as, the point at which transition from localized to spatiotemporal chaos 
occurs. We have checked it by looking at actual pictures for all values of $\tau$ 
including $\tau=0$.}

\bibitem[{\citenamefont{haye2}(2008)}]{haye2}
\bibinfo{author}{\bibfnamefont{J.} \bibnamefont{Fuchs}},
\bibinfo{author}{\bibfnamefont{J.} \bibnamefont{Schelter}},
\bibinfo{author}{\bibfnamefont{F.} \bibnamefont{Ginelli}},
\bibinfo{author}{\bibfnamefont{H.} \bibnamefont{Hinrichsen}},
\bibinfo{journal}{J .Stat. Mech }, \bibinfo{pages}{P04015} (\bibinfo{year}{2008}).

\bibitem[()]{footnote3}
\bibinfo{note}{Notice here that for $\tau>1$, the gap in the eigenvalue spectrum 
is not a well defined quantity. For e.g. let us consider the case $\tau=2$. 
Let $N=1$,  $2 \pi x_1(0)= \pi/2$, $\delta=1$, $k=1$ and $\epsilon=0$. The values 
of $\omega$, $x_1(1)$ and $x_1(2)$ can be arbitrary. Now the Jacobian at first time step is.
\begin{displaymath}
\left(\begin{array}{c c c}
0 & 0 & 1 \\
1 & 0 & 0 \\
0 & 1 & 0 \\
\end{array}\right),
\end{displaymath}
which clearly has complex eigenvalues. Thus it is not clear how one is
going to look at gaps in eigenvalue spectrum for $\tau>1$.
}

\bibitem[{\citenamefont{toeplitz}(2008)}]{toeplitz}
\bibinfo{author}{\bibfnamefont{W.} \bibnamefont{Yueh}},
\bibinfo{author}{\bibfnamefont{S. S.} \bibnamefont{Cheng,}}
  \bibinfo{journal}{ANZIAM J.} \textbf{\bibinfo{volume}{49}},
  \bibinfo{pages}{361-387} (\bibinfo{year}{2008}).

\bibitem[{\citenamefont{liuwang}(2009)}]{liuwang}
\bibinfo{author}{\bibfnamefont{D. Z.} \bibnamefont{Liu}},
\bibinfo{author}{\bibfnamefont{Z. D.} \bibnamefont{Wang}}
  \bibinfo{journal}{arXiv:0904.2958v2 [math.PR].} ,
 (\bibinfo{year}{2008}).
\bibitem[()]{footnote4}
\bibinfo{note}{We have not explicitly carried out the stability
analysis of synchronized state. However, it can be easily seen that
the Jacobian is a block-circulent matrix which can be
block-diagonalized to N blocks of size $(2 \times 2)$ which can easily be
diagonalized, see {\it e. g.}  P. M. Gade and R. E. Amritkar, Phys. Rev. A {\bf 47},
143 (1993). }



\bibitem[{\citenamefont{afraimovich}(1994)}]{afraimovich}
\bibinfo{author}{\bibfnamefont{V. S.}~\bibnamefont{Afraimovich,}} 
\bibinfo{author}{\bibfnamefont{V. I.}~\bibnamefont{Nekorkin}}, 
\bibinfo{author}{\bibfnamefont{G. V.}~\bibnamefont{Osipov}},
\bibinfo{author}{\bibfnamefont{V. D.}~\bibnamefont{Shalfeev}},
  \bibinfo{title}{ {\em{Stability, Structures and Chaos in Nonlinear Synchronization Networks}}},
  (\bibinfo{publisher}{World Scientific},
\bibinfo{address}{Singapore},\bibinfo{year}{1994})

\bibitem[{\citenamefont{kurths}(2002)}]{kurths}
\bibinfo{author}{\bibfnamefont{G. V.}~\bibnamefont{Osipov}}, 
\bibinfo{author}{\bibfnamefont{A. S.}~\bibnamefont{Pikovsky}},
\bibinfo{author}{\bibfnamefont{J.}~\bibnamefont{Kurths}},
  \bibinfo{journal}{Phys. Rev. Lett} \textbf{\bibinfo{volume}{88}},
  \bibinfo{pages}{054102} (\bibinfo{year}{2002}).

\bibitem[{\citenamefont{Sethia}(2008)}]{Sethia}
\bibinfo{author}{\bibfnamefont{G. C.} \bibnamefont{Sethia}},
\bibinfo{author}{\bibfnamefont{A.} \bibnamefont{Sen,}}
\bibinfo{author}{\bibfnamefont{F. M.} \bibnamefont{Atay}},
  \bibinfo{journal}{Phys. Rev. Lett} \textbf{\bibinfo{volume}{100}},
  \bibinfo{pages}{144102} (\bibinfo{year}{2008}).


\bibitem[{\citenamefont{kuramoto}(2002)}]{kuramoto}
\bibinfo{author}{\bibfnamefont{Y.}~\bibnamefont{Kuramoto}} \bibnamefont{and}
  \bibinfo{author}{\bibfnamefont{D.}~\bibnamefont{Battogtokh}},
  \bibinfo{journal}{Nonlinear Phen. in Complex Systems} \textbf{\bibinfo{volume}{5:4}},
  \bibinfo{pages}{380-385} (\bibinfo{year}{2002}).


\bibitem[{\citenamefont{abrams}(2004)}]{abrams}
\bibinfo{author}{\bibfnamefont{D. M.} \bibnamefont{Abrams}},
\bibinfo{author}{\bibfnamefont{S. H.} \bibnamefont{Strogatz,}}
  \bibinfo{journal}{Phys. Rev. Lett} \textbf{\bibinfo{volume}{93}},
  \bibinfo{pages}{174102} (\bibinfo{year}{2004}).
\bibitem[{\citenamefont{tass}(2008)}]{tass}
\bibinfo{author}{\bibfnamefont{O. E.} \bibnamefont{Omel'chenko}},
\bibinfo{author}{\bibfnamefont{Y. L.} \bibnamefont{Maestrenko,}}
\bibinfo{author}{\bibfnamefont{P. A.} \bibnamefont{tass}},
  \bibinfo{journal}{Phys. Rev. Lett} \textbf{\bibinfo{volume}{100}},
  \bibinfo{pages}{044105} (\bibinfo{year}{2008}).


\bibitem[{\citenamefont{sheeba}(2009)}]{sheeba}
\bibinfo{author}{\bibfnamefont{J. H.} \bibnamefont{Sheeba}},
\bibinfo{author}{\bibfnamefont{V. K.} \bibnamefont{Chandrasekar,}}
\bibinfo{author}{\bibfnamefont{M.} \bibnamefont{Lakshmanan}},
  \bibinfo{journal}{Phys. Rev. E} \textbf{\bibinfo{volume}{79}},
  \bibinfo{pages}{055203(R)} (\bibinfo{year}{2009}).
\bibitem[{\citenamefont{wiley}(2008)}]{wiley}
\bibinfo{author}{\bibfnamefont{D. M.} \bibnamefont{Abrams}},
\bibinfo{author}{\bibfnamefont{R.} \bibnamefont{Mirollo,}}
\bibinfo{author}{\bibfnamefont{S. H.} \bibnamefont{Strogatz}},
\bibinfo{author}{\bibfnamefont{D. A.} \bibnamefont{Wiley}},
  \bibinfo{journal}{Phys. Rev. Lett} \textbf{\bibinfo{volume}{101}},
  \bibinfo{pages}{084103} (\bibinfo{year}{2008}).

\bibitem[{\citenamefont{laing}(2009)}]{laing}
  \bibinfo{author}{\bibfnamefont{C. R.}~\bibnamefont{Laing}},
  \bibinfo{journal}{Chaos} \textbf{\bibinfo{volume}{19}},
  \bibinfo{pages}{013113} (\bibinfo{year}{2009}).

\end{thebibliography}
\end{document}